\documentclass[a4paper,UKenglish,cleveref,autoref,thm-restate]{lipics-v2021}

\usepackage{semantic}
\usepackage{mathpartir}
\usepackage{float}
\usepackage[normalem]{ulem}
\usepackage{stmaryrd}
\usepackage{enumitem}

\usepackage{tikz}
\usetikzlibrary{arrows.meta}
\usetikzlibrary{positioning}
\usetikzlibrary{intersections}
\usetikzlibrary{shapes.geometric}
\usetikzlibrary{shapes.multipart}
\usetikzlibrary{calc}
\usetikzlibrary{fit}
\usetikzlibrary{backgrounds}
\usetikzlibrary{decorations.pathreplacing}

\usepackage{listings-rust}
\usepackage{wrapfig}

\definecolor{Cerulean}{rgb}{0.20,0.31,0.0.90}
\definecolor{RedOrange}{rgb}{0.70,0.59,0.0.20}
\definecolor{ForestGreen}{rgb}{0.1,0.60,0.2}
\definecolor{SomeRed}{rgb}{0.80,0.3,0.0.20}
\definecolor{lightGray}{rgb}{0.85,0.85,0.86} %
\definecolor{lGray}{rgb}{0.31,0.31,0.33} %
\definecolor{lYellow}{rgb}{0.99,0.78,0.07} %
\definecolor{lila}{rgb}{0.7,0.30,0.77} %
\definecolor{cloudyblue}{rgb}{206,221,246}

\usepackage{ifthen}

\newboolean{showcomments}
\setboolean{showcomments}{false}

\makeatletter
\newcommand{\mynote}[3]{%
  \ifthenelse{\boolean{showcomments}}{%
   \fbox{\bfseries\sffamily\scriptsize#1}%
   {\small\textsf{\emph{\color{#3}{#2}}}}}%
  {%
   \@bsphack
   \@esphack
  }%
}
\makeatother

\definecolor{asparagus}{rgb}{0.53, 0.66, 0.42}

\newcommand{\out}[1]{}

\setboolean{showcomments}{false}

\usepackage{xspace}

\newcommand{\itree}{\text{ITree}\xspace}
\newcommand{\itrees}{\text{ITrees}\xspace}

\newcommand{\WP}[2]{\texttt{WP}~#1~\lbrace #2 \rbrace}

\newcommand{\updateMod}{\mid \! \Rrightarrow}

\newcommand{\itreeER}[2]{\ensuremath{\texttt{itree}\;{#1}\;{#2}}}
\newcommand{\Ret}[1]{\ensuremath{\texttt{Ret}(#1)}}
\newcommand{\Tau}[1]{\ensuremath{\texttt{Tau}(#1)}}
\newcommand{\VisA}[3]{\ensuremath{\texttt{Vis}_{#1}(#2, #3)}}

\newcommand{\wpi}[3]{\ensuremath{ \mathrm{wpi}_{#1}\;#2\;\{ #3 \} }}

\setboolean{showcomments}{true} 

\newcommand{\WPt}[2]{\mathrm{WP}_{\mathsf{t}}\;#1\;\lbrace #2\rbrace}
\newcommand{\WPsrc}[2]{\mathrm{WP}_{\mathit{src}}\;#1\;\lbrace #2\rbrace}
\newcommand{\WPsrctot}[2]{\mathrm{WP}^{\mathit{tot}}_{\mathit{src}}\;#1\;\lbrace #2\rbrace}
\newcommand{\WPg}[2]{\mathrm{WP}_{\mathit{graph}}\;#1\;\lbrace #2\rbrace}
\newcommand{\WPx}[3]{\mathrm{WP}_{\mathit{#1}}\;#2\;\lbrace #3\rbrace}

\title{From Sequential to Dataflow: Formally-verified Compilation for Concurrent and Distributed Architectures}
\titlerunning{From Sequential to Dataflow}

\ifx\authoranonymous\relax
\author{Anonymous}{Anonymous}{}{}{}
\else
\author{Sebastian Ertel}{Barkhausen Institut, Dresden, Germany}{sebastian.ertel@barkhauseninstitut.org}{}{}
\author{Mohamed Elsheikh}{Barkhausen Institut, Dresden, Germany \and Technische Universit\"at Dresden, Germany}{mohamed\_abdelhameed.elsheikh@mailbox.tu-dresden.de}{}{}
\author{Carmine Abate}{Barkhausen Institut, Dresden, Germany}{carmine.abate@barkhauseninstitut.org}{}{}
\fi
\authorrunning{S. Ertel, M. Elsheikh, and C. Abate}
\Copyright{Sebastian Ertel, Mohamed Elsheikh, and Carmine Abate}



\acknowledgements{This work has been supported by funding from the Agentur f\"ur Innovation in der
Cybersicherheit GmbH (Cyberagentur).}

\hideLIPIcs%

\begin{document}
\nolinenumbers

\maketitle

\begin{abstract}
On serverless cloud platforms, an application consists of many functions that communicate through the
network, which makes it hard to debug, monitor and verify.
A growing line of work therefore lets developers write a sequential, imperative program and compiles it
into a distributed application.
The correctness of these compilers, however, rests on informal arguments, and a mechanized guarantee is
hard to obtain: the compiled program runs concurrently but must keep the order of the updates to each
state, it must not deadlock, and it may be run by many different executors.

We implement a compiler from a sequential, imperative language with state, loops and tail recursion to a
dataflow language, and verify it in Rocq.
The compiler translates a program directly into a graph of nodes connected by channels: each state
becomes a chain of channels that orders its updates, and each tail recursion becomes a loop in the graph
that feeds the result of one round back as the argument of the next.
To verify the compiler, we give both languages a
semantics with interaction trees, derive weakest precondition calculi from
these semantics, and show that every weakest precondition
proved for the sequential program holds for its compiled graph.
Our result is stated against a contract of graph execution, which fixes the order of calls and
iterations, leaves every other firing free, and requires that the graph never gets stuck.
It therefore holds for every executor that meets the contract. We show that a scheduler meets it under
every schedule, and so does an executor without a scheduler, in which nodes fire in parallel as soon as
their inputs hold data.
\end{abstract}

\raggedbottom
\definecolor{fData}{rgb}{0.15,0.15,0.18}
\definecolor{fState}{rgb}{0.85,0.42,0.05}
\definecolor{fFront}{rgb}{0.16,0.38,0.75}
\definecolor{fLoop}{rgb}{0.90,0.95,1.00}
\definecolor{fInit}{rgb}{0.95,0.95,0.95}
\definecolor{fNode}{rgb}{1.00,1.00,1.00}
\definecolor{fProof}{rgb}{0.20,0.55,0.25}

\tikzset{
  fnode/.style={draw,rounded corners=2pt,fill=fNode,font=\scriptsize\sffamily,
                inner sep=2.5pt,minimum height=5mm,align=center},
  fdata/.style={->,>=Latex,thick,draw=fData},
  fstate/.style={->,>=Latex,thick,dashed,draw=fState},
  fdang/.style={->,>=Latex,thin,draw=fData!50},
  ffront/.style={densely dashed,very thick,draw=fFront},
  flab/.style={font=\scriptsize,inner sep=1pt},
  fregion/.style={rounded corners=4pt,inner sep=4pt},
  frowA/.style={draw,fill=fFront!15,font=\tiny,inner sep=1pt,minimum height=3.6mm,minimum width=5mm},
  frowB/.style={draw,fill=fState!22,font=\tiny,inner sep=1pt,minimum height=3.6mm,minimum width=5mm},
  fstep/.style={-{Implies},double,double distance=1.6pt,thick,draw=fProof},
  fcell/.style={draw,fill=white,font=\tiny,inner sep=1pt,minimum height=3.6mm,minimum width=5mm},
}

\newcommand{\evS}{\Downarrow_S}
\newcommand{\evRows}{\Downarrow^{\mathrm{rows}}_S}
\newcommand{\evBatch}{\Downarrow^{\mathrm{batch}}_S}
\newcommand{\evGraph}{\Downarrow^{G}_S}
\newcommand{\divS}{\Uparrow_S}
\newcommand{\divBatch}{\Uparrow^{\mathrm{batch}}_S}
\newcommand{\divGraph}{\Uparrow^{G}_S}
\newcommand{\stepM}{\longrightarrow_S}
\newcommand{\stepsM}{\longrightarrow^{*}_S}
\newcommand{\compiles}[4]{#1\vdash #2\rightsquigarrow(#3)}
\newcommand{\realizes}{\Vdash}
\newcommand{\wireOf}{\mathsf{wire}}

\newcommand{\rowA}[1]{{\setlength{\fboxsep}{1pt}\fcolorbox{fData}{fFront!15}{$#1$}}}
\newcommand{\rowB}[1]{{\setlength{\fboxsep}{1pt}\fcolorbox{fData}{fState!22}{$#1$}}}

\newcommand{\condrust}{ConDRust\xspace}
\newcommand{\verifiedclaim}[1]{\textcolor{fProof!80!black}{#1}}

\newcommand{\srclang}{SI$_0$\xspace}

\newcommand{\sletstate}{\ensuremath{\mathbf{letstate}}\xspace}
\newcommand{\sfor}{\ensuremath{\mathbf{for}}\xspace}
\newcommand{\slet}{\ensuremath{\mathbf{let}}\xspace}
\newcommand{\strfix}{\ensuremath{\mathbf{trfix}}\xspace}

\section{Introduction}\label{sec:intro}

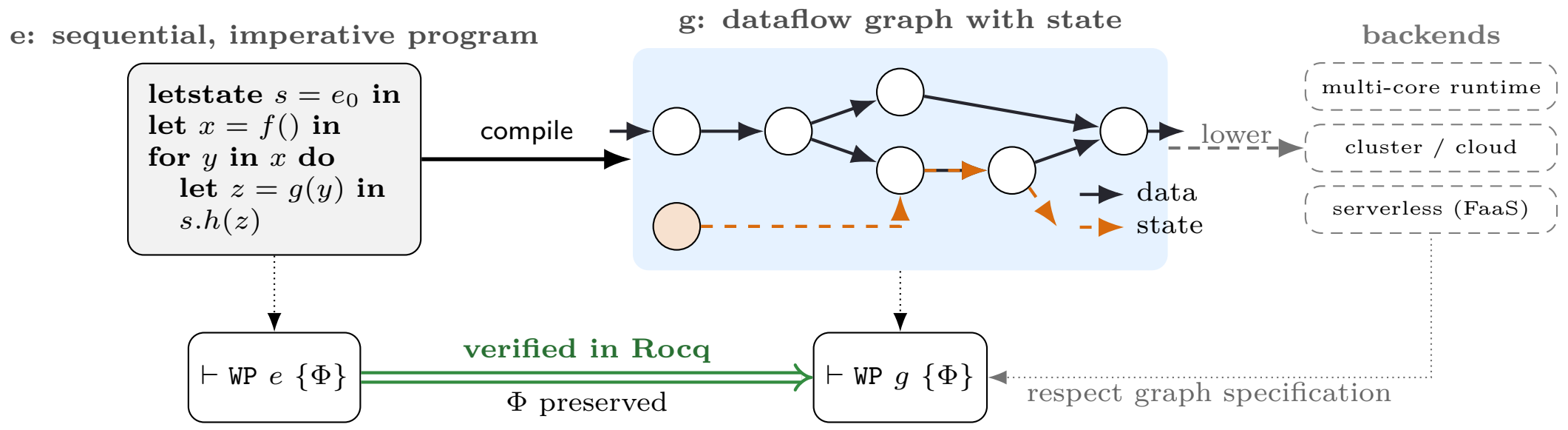
\begin{figure}[t]
\centering
\begin{tikzpicture}[x=1cm,y=1cm,
    side/.style={font=\scriptsize\bfseries,text=black!70},
    prog/.style={draw,rounded corners,fill=fInit,align=left,font=\scriptsize,inner sep=5pt},
    sem/.style={draw,rounded corners,align=center,font=\scriptsize,inner sep=3pt,minimum height=0.8cm},
    an/.style={draw,circle,fill=white,minimum size=4.2mm,inner sep=0pt},
    big/.style={-{Latex[length=2.5mm]},line width=1pt},
    fut/.style={draw=black!50,densely dashed,rounded corners,font=\tiny,minimum width=2.25cm,minimum height=0.42cm,inner sep=1pt}]

  \node[prog] (src) at (2.0,1.1) {%
    $\mathbf{letstate}\ s=e_0\ \mathbf{in}$\\
    $\mathbf{let}\ x=f()\ \mathbf{in}$\\
    $\mathbf{for}\ y\ \mathbf{in}\ x\ \mathbf{do}$\\
    $\quad\mathbf{let}\ z=g(y)\ \mathbf{in}$\\
    $\quad s.h(z)$};
  \node[side,anchor=south] at (src.north) {e: sequential, imperative program};
  \node[an] (n1) at (5.6,1.35) {};
  \node[an] (n2) at (6.6,1.35) {};
  \node[an] (n3) at (7.6,1.7) {};
  \node[an] (n4) at (7.6,1.0) {};
  \node[an] (n5) at (8.6,1.0) {};
  \node[an] (n6) at (9.6,1.35) {};
  \node[an,fill=fState!20] (s0) at (5.6,0.5) {};
  \draw[fdata] (5.0,1.35) -- (n1);
  \draw[fdata] (n1) -- (n2);
  \draw[fdata] (n2) -- (n3);
  \draw[fdata] (n2) -- (n4);
  \draw[fdata] (n4) -- (n5);
  \draw[fdata] (n3) -- (n6);
  \draw[fdata] (n5) -- (n6);
  \draw[fdata] (n6) -- (10.15,1.35);
  \draw[fstate] (s0) -| (n4);
  \draw[fstate] (n4) -- (n5);
  \draw[fstate] (n5.south east) -- ++(0.25,-0.35);
  \node[flab,anchor=west] at (9.15,0.8) {\tikz[baseline=-0.5ex]\draw[fdata](0,0)--(0.4,0); data};
  \node[flab,anchor=west] at (9.15,0.5) {\tikz[baseline=-0.5ex]\draw[fstate](0,0)--(0.4,0); state};
  \begin{scope}[on background layer]
    \node[fregion,fill=fLoop,fit=(n1)(n3)(n6)(s0),inner sep=5pt] (df) {};
  \end{scope}
  \node[side,anchor=south] at (df.north) {g: dataflow graph with state};
  \node[fut] (b1) at (12.35,1.75) {multi-core runtime};
  \node[fut] (b2) at (12.35,1.2) {cluster / cloud};
  \node[fut] (b3) at (12.35,0.65) {serverless (FaaS)};
  \node[fit=(b1)(b3),inner sep=1pt] (be) {};
  \node[side,anchor=south,text=black!55] at (be.north) {backends};
  \draw[-{Latex},thick,densely dashed,draw=black!55] (df.east |- b2) -- node[flab,above,text=black!60]{lower} (b2.west);
  \draw[big] (src.east) -- node[font=\scriptsize,above,align=center]{$\mathsf{compile}$} (df.west |- src.east);
  \node[sem] (ss) at (2.0,-0.85) {$\vdash\WP{e}{\Phi}$};
  \node[sem] (fs) at (7.6,-0.85) {$\vdash\WP{g}{\Phi}$};
  \draw[-{Latex},densely dotted] (src.south) -- (ss.north);
  \draw[-{Latex},densely dotted] (df.south -| fs.north) -- (fs.north);
  \draw[-{Latex},densely dotted,draw=black!55] (be.south) |- node[flab,below,pos=0.75,text=black!60]{respect graph specification} (fs.east);
  \draw[-{Implies},double,double distance=1.6pt,thick,draw=fProof] (ss.east) --
    node[font=\scriptsize,above,align=center,text=fProof!80!black]{\textbf{verified in Rocq}}
    node[font=\scriptsize,below,align=center]{$\Phi$ preserved}
    (fs.west);
\end{tikzpicture}
\caption{A sequential, imperative program with state and control flow $e$ (top left)
is compiled into a dataflow graph $g$ whose edges carry data and the ordering of state accesses (top middle).
A machine-checked result guarantees that compilation preserves specifications proven
in a weakest precondition calculus for the source language (bottom): 
every specification $\Phi$ proved for $e$ also holds for the graph $g$, whether a
scheduler executes it, under any schedule, or its nodes fire in parallel, without one.
The result holds for every executor that respects the graph specification --
calls and iterations follow the source order, while all other firings run freely,
without deadlock -- so future backends to multi-core, cluster and serverless cloud
targets (right, dashed) inherit it.}
\label{fig:intro-goal}
\end{figure}

Cloud computing~\cite{berkeley-cloud} is the predominant execution paradigm for large-scale applications.
In its serverless form~\cite{berkeley-serverless}, the provider runs code without developers having to
provision or maintain servers: resources scale automatically on demand, and billing is based on actual
usage rather than on pre-allocated capacity.
The most fine-grained form of serverless computing is Function as a Service
(FaaS)~\cite{aws_lambda, azure_functions, google_cloud_functions, ibm_cloud_functions,
cloudflare_workers, oracle_functions, alibaba_function_compute}, in which the provider runs individual
functions.

In order to scale and tolerate failures, the cloud runs functions in parallel across the cores, 
racks and nodes of a cluster and restarts failing functions on another machine.
Scalability and fault tolerance both rest on a distributed, seemingly stateless execution model that imposes, however, 
a change also in the programming model.
An application that is conceptually one sequential program must be implemented as many functions, connected through the
network instead of through function calls, and is therefore much harder to debug, monitor and verify.

A growing line of work therefore lets developers write stateful programs in a sequential, imperative
language, and compiles them into distributed applications~\cite{cascade, cryptosyn, viaduct, monolift,
condrust}.
Cascade~\cite{cascade} transforms imperative microservices into stateful dataflow systems, for performance
and fault tolerance.
ConDRust~\cite{condrust} compiles verified sequential Rust programs into deterministic concurrent
implementations.
Viaduct~\cite{viaduct} compiles programs annotated with security policies into secure distributed
implementations that use cryptographic mechanisms.
Compilation from sequential programs to dataflow also targets hardware, such as coarse-grained
reconfigurable arrays and FPGAs, and there its correctness has been mechanized.
FlowCert~\cite{flowcert} validates each run of a compiler from LLVM to dataflow graphs, and shows that the
graph reaches the same final memory as the source under every schedule; Lin et al.~\cite{letitflow}
verify the core passes of such a compiler in Lean, by a forward simulation and determinacy of the graph.
These systems come from far-apart communities, but they make the same promise: developers reason about a
sequential program, and the compiler takes care of parallel or distributed execution.

For distributed and serverless targets, this promise rests on informal arguments: ConDRust argues that its
compilation preserves semantics, and Acay et al.~\cite{cryptosyn} prove on paper, for a core of Viaduct's
language without loops, that its compilation preserves security.
The mechanized results for hardware do not carry over, for three reasons.
First, they relate the final memories of the source and of the graph, which describes a closed program; a
serverless application instead calls functions and services known only through their specifications, and
what its developers verify is a specification, so the guarantee must carry specifications.
Second, the functions of a serverless application share no memory: the nodes of the compiled program fire in
an order the source does not fix, the updates of each state must keep the source's order without a memory
that arbitrates them, and the program must not deadlock.
Third, a hardware target has one execution model, while distributed execution is not one fixed machine,
so the guarantee should hold for every executor, including backends that do not exist yet.

In this work, we implement a compiler from a sequential, imperative language to a dataflow
language and verify it in the Rocq proof assistant. \Cref{fig:intro-goal} depicts the guarantees given by our approach.
We show that the compiler preserves specifications: every weakest precondition proved for the sequential program holds for its compiled graph.
The key to this result is a specification of graph execution, which any executor
has to respect: calls and iterations happen in the order of the sequential program, every other firing runs
concurrently, and the graph never gets stuck.

\smallskip
More precisely, \emph{our contributions} are:
\begin{enumerate}
  \item We give two languages a semantics with interaction trees: a sequential, imperative language with
        state, loops over lists and tail recursion, and a dataflow language, whose programs
        build a graph of nodes connected by channels and then execute it (\Cref{sec:source,sec:target}).
        The execution of a graph is a single event, so the same semantics covers a scheduler,
        which fires one node at a time, and an executor without a scheduler, in which nodes fire in parallel
        as soon as their inputs hold data.
        Following Vistrup et al.~\cite{plalacarte}, we derive sound weakest precondition calculi for both
        languages from their semantics, by giving each event a logical interpretation.
  \item We present a compiler from the sequential language to the dataflow language that needs no
        compiler-specific intermediate representation: it translates expressions directly into graphs,
        turns each state into a chain of channels that orders its updates, and each tail recursion into a
        loop in the graph that feeds the result of one round back as the argument of the next.
        The compiled graphs expose two kinds of parallelism: calls that neither use each other's results
        nor share a state run in parallel, and the iterations of a \sfor loop overlap in a pipeline, as
        every node of the body handles one element after another. Calls on one state keep their order,
        and the rounds of a tail recursion, each of which needs the result of the previous one, run one
        after another.
        It succeeds on every well-formed program, and every channel of a compiled graph has one writer and
        at most one reader (\Cref{sec:compiler-impl}).
  \item We define $\mathrm{WP}_{\mathsf{t}}$, a specification of graph execution that fixes the order of
        calls and iterations, leaves every other firing free, and requires that the graph never gets stuck.
        We prove that the source-order scheduler, and the demonic scheduler under every schedule, implement
        $\mathrm{WP}_{\mathsf{t}}$ for every graph (\Cref{sec:verification}).
  \item We formalize both languages, their program logics and the compiler in Rocq, and prove that the
        compiler preserves weakest preconditions: every specification proved for a sequential program
        holds for its compiled graph, in source order for every handler of iterations, and under every
        schedule and under parallel execution without a scheduler when iterations are unguarded
        (\Cref{sec:verification}).
        We apply the result to Labyrinth, a routing algorithm from the STAMP benchmark
        suite~\cite{stamp} that connects pairs of endpoints on a shared grid by routes that must not
        overlap: the sequential proof that its routes are disjoint carries over to its compiled graph, run
        in parallel (\Cref{sec:overview}).
        To ease navigation of the formalization, a web page, linked from the README of the artifact,
        mirrors the definitions and results of the paper and links each of them to its Rocq source.
\end{enumerate}
Since the result holds for every executor that implements $\mathrm{WP}_{\mathsf{t}}$, future backends for
multi-core, cluster and serverless targets inherit it.

\smallskip
The rest of the paper is organized as follows.
\Cref{sec:background} recalls interaction trees and how program logics are obtained from \itree
semantics.
\Cref{sec:overview} introduces the approach on Labyrinth: its
sequential implementation and verification, the compiled dataflow graph, and why the graph inherits the
program's specification.
\Cref{sec:source} presents the source language, its semantics and its program logic.
\Cref{sec:target} presents the dataflow language, its semantics, and its executors, with and
without a scheduler.
\Cref{sec:compiler} presents the compiler and proves that it preserves weakest preconditions.
\Cref{sec:relatedW} discusses related work, and \Cref{sec:conclusions} concludes.

\section{Background on Interaction Trees}\label{sec:background}

Interaction trees (\itrees) are a coinductive representation of recursive,
effectful computations in Rocq~\cite{itrees}.  An \itree of type
\itreeER{E}{R} is parameterized by an event signature $E$ and a return
type $R$.  It has three constructors: \Ret{r} terminates and returns $r : R$; \Tau{t} takes an
internal, unobservable step and continues as $t$; and \VisA{A}{\epsilon}{k} emits an event
$\epsilon : E\ A$ and, after receiving an answer $a : A$ from the environment, continues as $k\,a$.
Because the datatype is coinductive, its branches may be infinite.  It can
therefore represent both reactive programs that produce unbounded sequences of
events and programs that diverge silently through an infinite sequence of
$\mathsf{Tau}$ steps.

\emph{Algebraic structure.} The main equivalence on \itrees is \emph{weak bisimulation}, or \emph{equivalence up to tau}, written
$\approx$: it relates trees that return equal values and emit equal events, ignoring finite numbers of
silent steps, while still distinguishing silent divergence from termination.
For every event signature $E$, \itreeER{E}{R} forms a monad whose unit is
$\mathsf{Ret}$ and whose bind $\mathsf{bind}\ t\ k$, written $x\leftarrow t;\,k\,x$, replaces each
return leaf \Ret{r} of $t$ with the continuation $k\,r$, while preserving silent steps and visible events.
Monadic laws hold up to weak bisimulation.
Event signatures combine by disjoint sum: a tree of type \itreeER{(E_1+E_2)}{R} may emit the events of
both.
\itrees also come with an iteration combinator, $\mathsf{loop}$ (\texttt{ITree.iter}~\cite{itrees}),
which implements loops: its body returns either $\mathsf{Continue}(a)$, to run again from $a$, or
$\mathsf{Done}(b)$, to end the loop with $b$; a loop that never returns $\mathsf{Done}$ is an infinite tree.

\emph{Computational and logical interpretations.} Events are initially uninterpreted: an event signature describes the
interface between a computation and its environment, and a handler assigns
meaning to it. A \emph{computational handler} maps each event to a
computation; an interpreter folds it over a tree, mapping $\mathsf{Ret}$ to
monadic return and commuting with bind, so that interpreters preserve the
monadic structure and the equational theory. A \emph{logical handler} $H$
instead assigns each event a specification~\cite{plalacarte}: $H(\epsilon,\Psi)$ is what must hold
before the event $\epsilon$ so that the program can continue with the postcondition $\Psi$ on the
answer; a weakest-precondition interpretation
lifts it to an entire \itree, mapping $\mathsf{Ret}$ to the postcondition
and composing through bind, so that an \itree semantics gives rise
compositionally to a program logic for the represented language. Both
computational and logical handlers for individual effects can be combined
over disjoint sums of event signatures: $H_1\oplus H_2$ handles the events of $E_1+E_2$, each with its own
handler.

\emph{Program logics from \itree semantics.} Vistrup et~al.~\cite{plalacarte} derive a program logic for a language from its \itree semantics: a
denotation of programs into \itreeER{E}{R} and a logical handler for each event in $E$.
The foundation is a general weakest-precondition calculus for \itrees,
\[
\wpi{H}{t}{\Phi} \mathrel{:=} \updateMod
\begin{cases}
  \Phi(r)
    & \text{if } t = \Ret{r}, \\
      \wpi{H}{t'}{\Phi}
    & \text{if } t = \Tau{t'}, \\
  H\bigl(\epsilon, \lambda a.\,\wpi{H}{k(a)}{\Phi}\bigr)
    & \text{if } t = \VisA{A}{\epsilon}{k}.
\end{cases}
\]
Thus, a return establishes the postcondition, a silent step continues with the
weakest precondition of its continuation, and a visible event delegates reasoning to
the logical handler.
The definition of $\mathrm{wpi}$ is a least fixpoint, so a weakest precondition rules out infinite runs of
the tree. To reason about infinite \itrees, a handler can guard the continuation of some events, for
instance an event that marks a step of computation, by a later modality~$\triangleright$.
The outer update modality permits updates to Iris~\cite{iris} ghost
state during each step.

\section{Overview}\label{sec:overview}

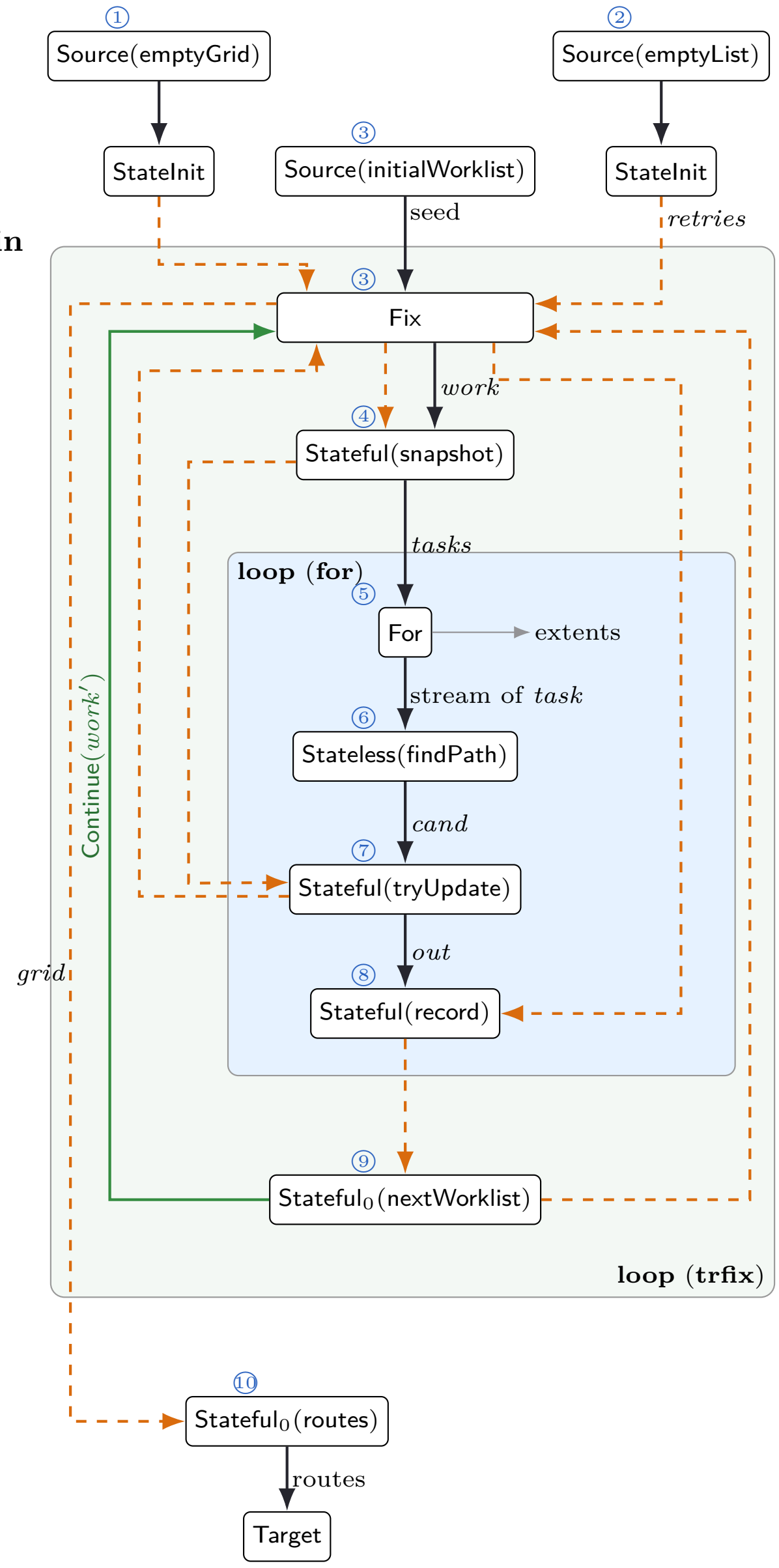
\begin{figure}[p]
\centering
\begin{minipage}[c]{0.40\textwidth}
\begin{subfigure}{\linewidth}
\small
\[
\begin{array}{@{}r@{\ \ }l@{}}
\text{\scriptsize\textcircled{\tiny 1}} & \mathbf{letstate}\ \mathit{grid}=\mathsf{emptyGrid}()\ \mathbf{in}\\[2pt]
\text{\scriptsize\textcircled{\tiny 2}} & \mathbf{letstate}\ \mathit{retries}=\mathsf{emptyList}()\ \mathbf{in}\\[2pt]
\text{\scriptsize\textcircled{\tiny 3}} & \mathbf{let}\ \_=\mathbf{trfix}\ \mathit{work}=\mathsf{initialWorklist}()\ \mathbf{in}\\[2pt]
\text{\scriptsize\textcircled{\tiny 4}} & \quad\mathbf{let}\ \mathit{tasks}=\mathit{grid}.\mathsf{snapshot}(\mathit{work})\ \mathbf{in}\\[2pt]
\text{\scriptsize\textcircled{\tiny 5}} & \quad\mathbf{let}\ \_=\mathbf{for}\ \mathit{task}\ \mathbf{in}\ \mathit{tasks}\ \mathbf{do}\\[2pt]
\text{\scriptsize\textcircled{\tiny 6}} & \quad\quad\mathbf{let}\ \mathit{cand}=\mathsf{findPath}(\mathit{task})\ \mathbf{in}\\[2pt]
\text{\scriptsize\textcircled{\tiny 7}} & \quad\quad\mathbf{let}\ \mathit{out}=\mathit{grid}.\mathsf{tryUpdate}(\mathit{cand})\ \mathbf{in}\\[2pt]
\text{\scriptsize\textcircled{\tiny 8}} & \quad\quad\mathit{retries}.\mathsf{record}(\mathit{out})\\[2pt]
\text{\scriptsize\textcircled{\tiny 9}} & \quad\mathbf{in}\ \mathit{retries}.\mathsf{nextWorklist}()\\[2pt]
\text{\scriptsize\textcircled{\tiny 10}} & \mathbf{in}\ \mathit{grid}.\mathsf{routes}()
\end{array}
\]
\caption{Sequential program.}\label{fig:overview-labyrinth-src}
\end{subfigure}

\vspace{2em}
\addtocounter{subfigure}{1}%
\begin{subfigure}{\linewidth}
\centering
\begin{tikzpicture}[x=0.4cm,y=0.4cm,
    gridc/.style={draw=black!25,very thin},
    endp/.style={draw,minimum size=0.36cm,inner sep=0pt,font=\tiny},
    pA/.style={draw=fFront,line width=1.6pt,line cap=round,line join=round},
    pB/.style={draw=fState,line width=1.6pt,line cap=round,line join=round},
    claimA/.style={fill=fFront!30},
    claimB/.style={fill=fState!35},
    desc/.style={font=\scriptsize,anchor=west,align=left,text width=2.45cm}]
  \newcommand{\lgrid}{\draw[gridc] (0,0) grid (7,5);}
  \newcommand{\lends}{%
    \node[endp,fill=fFront!15] at (1.5,1.5) {$p_0$};
    \node[endp,fill=fFront!15] at (5.5,1.5) {$p_0$};
    \node[endp,fill=fState!20] at (3.5,4.5) {$p_1$};
    \node[endp,fill=fState!20] at (3.5,0.5) {$p_1$};}
  \begin{scope}[yshift=0cm]
    \lgrid
    \draw[pA] (1.5,1.5) -- (5.5,1.5);
    \draw[pB] (3.5,4.5) -- (3.5,0.5);
    \node[font=\tiny,text=fFront,anchor=north] at (2.5,1.4) {$r_0$};
    \node[font=\tiny,text=fState!80!black,anchor=west] at (3.6,3.0) {$r_1$};
    \lends
    \node[desc] at (7.5,2.5) {\textbf{1.} Round 0, search on snapshot $G_0$: the two searches are independent.};
  \end{scope}
  \begin{scope}[yshift=-2.7cm]
    \fill[claimA] (1,1) rectangle (6,2);
    \lgrid
    \draw[pA] (1.5,1.5) -- (5.5,1.5);
    \draw[pB,densely dotted,draw=fState!60] (3.5,4.5) -- (3.5,0.5);
    \node[font=\small,text=red!70!black] at (3.5,1.5) {$\boldsymbol\times$};
    \lends
    \node[desc] at (7.5,2.5) {\textbf{2.} Round 0, commit in worklist order: $r_0$ commits ($G_1$); $r_1$ crosses it ($\times$), so $p_1$ is retried.};
  \end{scope}
  \begin{scope}[yshift=-5.4cm]
    \fill[claimA] (1,1) rectangle (6,2);
    \fill[claimB] (3,4) rectangle (7,5) (6,0) rectangle (7,4) (3,0) rectangle (6,1);
    \lgrid
    \draw[pB] (3.5,4.5) -- (6.5,4.5) -- (6.5,0.5) -- (3.5,0.5);
    \node[font=\tiny,text=fState!80!black,anchor=east] at (6.4,2.6) {$r_2$};
    \lends
    \node[desc] at (7.5,2.5) {\textbf{3.} Round 1, search on $G_1$ and commit: $p_1$ takes the detour $r_2$ ($G_2$).};
  \end{scope}
\end{tikzpicture}
\caption{Rounds on two pairs.}\label{fig:labyrinth-viz}
\end{subfigure}
\end{minipage}\hfill
\setcounter{subfigure}{1}%
\begin{subfigure}[c]{0.58\textwidth}
\centering
\begin{tikzpicture}[x=1cm,y=0.9cm,
    ctl/.style={-{Latex},thick,draw=fProof},
    tag/.style={label={[flab,text=fFront,inner sep=0.5pt]above left:\textcircled{\tiny #1}}}]
  \node[fnode,tag=1] (eg) at (0.3,0) {$\mathsf{Source}(\mathsf{emptyGrid})$};
  \node[fnode] (ig) at (0.3,-1.3) {$\mathsf{StateInit}$};
  \node[fnode,tag=2] (el) at (5.4,0) {$\mathsf{Source}(\mathsf{emptyList})$};
  \node[fnode] (ir) at (5.4,-1.3) {$\mathsf{StateInit}$};
  \node[fnode,tag=3] (iw) at (2.8,-1.3) {$\mathsf{Source}(\mathsf{initialWorklist})$};
  \draw[fdata] (el) -- (ir);
  \draw[fdata] (eg) -- (ig);
  \begin{scope}[on background layer]
    \draw[rounded corners=4pt,fill=fProof!6,draw=black!40] (-0.8,-2.15) rectangle (6.55,-14.0);
    \draw[rounded corners=3pt,fill=fLoop,draw=black!40] (1.0,-5.6) rectangle (6.15,-11.5);
  \end{scope}
  \node[flab,anchor=south east] at (6.5,-13.95) {$\mathbf{loop}$ (\textbf{trfix})};
  \node[flab,anchor=north west] at (1.05,-5.65) {$\mathbf{loop}$ (\textbf{for})};
  \node[fnode,tag=3,minimum width=2.6cm] (fx) at (2.8,-2.95) {$\mathsf{Fix}$};
  \node[fnode,tag=4] (sn) at (2.8,-4.5) {$\mathsf{Stateful}(\mathsf{snapshot})$};
  \node[fnode,tag=5] (fo) at (2.8,-6.5) {$\mathsf{For}$};
  \node[fnode,tag=6] (fp) at (2.8,-7.9) {$\mathsf{Stateless}(\mathsf{findPath})$};
  \node[fnode,tag=7] (tu) at (2.8,-9.4) {$\mathsf{Stateful}(\mathsf{tryUpdate})$};
  \node[fnode,tag=8] (rc) at (2.8,-10.8) {$\mathsf{Stateful}(\mathsf{record})$};
  \node[fnode,tag=9] (nw) at (2.8,-12.9) {$\mathsf{Stateful}_0(\mathsf{nextWorklist})$};
  \node[fnode,tag=10] (ro) at (1.6,-15.4) {$\mathsf{Stateful}_0(\mathsf{routes})$};
  \node[fnode] (tg) at (1.6,-16.7) {$\mathsf{Target}$};
  \draw[fdata] (iw) -- node[flab,right,pos=0.15]{seed} (fx);
  \draw[fdata] ([xshift=0.3cm]fx.south) -- node[flab,right]{$\mathit{work}$} ([xshift=0.3cm]sn.north);
  \draw[fdata] (sn) -- node[flab,right]{$\mathit{tasks}$} (fo);
  \draw[fdata] (fo) -- node[flab,right]{stream of $\mathit{task}$} (fp);
  \draw[fdata] (fp) -- node[flab,right]{$\mathit{cand}$} (tu);
  \draw[fdata] (tu) -- node[flab,right]{$\mathit{out}$} (rc);
  \draw[fdang] (fo.east) -- ++(1.0,0) node[flab,right]{extents};
  \draw[fdata] (ro) -- node[flab,right]{routes} (tg);
  \coordinate (fxw) at ([yshift=-4pt]fx.west);
  \coordinate (fxe) at ([yshift=-4pt]fx.east);
  \draw[ctl] (nw.west) -- (-0.2,-12.9) -- node[flab,sloped,above,text=fProof!80!black,pos=0.5]
     {$\mathsf{Continue}(\mathit{work}')$} (-0.2,0 |- fxw) -- (fxw);
  \draw[fstate] (ig) -- (0.3,-2.35) -| ([xshift=-1.0cm]fx.north);
  \draw[fstate] ([xshift=-0.2cm]fx.south) -- ([xshift=-0.2cm]sn.north);
  \draw[fstate] ([yshift=-2pt]sn.west) -| (0.6,-9.33) -- ([yshift=2pt]tu.west);
  \draw[fstate] ([yshift=-2pt]tu.west) -| (0.1,-3.55) -| ([xshift=-0.9cm]fx.south);
  \draw[fstate] ([yshift=4pt]fx.west) -| node[flab,left,pos=0.8]{$\mathit{grid}$} (-0.6,-15.4) -- (ro.west);
  \draw[fstate] (ir) |- node[flab,right,pos=0.1]{$\mathit{retries}$} ([yshift=4pt]fx.east);
  \draw[fstate] ([xshift=0.9cm]fx.south) |- (5.6,-3.65) |- (rc.east);
  \draw[fstate] (rc) -- (nw);
  \draw[fstate] (nw.east) -| (6.3,0 |- fxe) -- (fxe);
\end{tikzpicture}
\caption{Compiled dataflow graph.}\label{fig:overview-labyrinth-dfg}
\end{subfigure}
\caption{Labyrinth: the sequential program (a), its compiled dataflow graph (b), and the algorithm's rounds on
two pairs $p_0$ (blue) and $p_1$ (orange) (c). In (b), circled tags: the source line each node comes from. Solid: data; dashed: state. The \textbf{for} becomes a loop group processing one stream element
per item; each state becomes a chain of tokens; the \textbf{trfix} becomes a loop group headed by a $\mathsf{Fix}$
node, with back edges for the round's result ($\mathsf{Continue}$) and for the states' tokens, which return to
$\mathsf{Fix}$ at the end of every round.
Omitted: the trigger stream ($\mathsf{Const}$, $\mathsf{Fanout}$), the copies of the round's argument
($\mathsf{Fanout}$), the loop's unit value ($\mathsf{Value}$), and the outputs of $\mathsf{Fix}$ that Labyrinth
does not use (the result of the \textbf{trfix} and the final token of $\mathit{retries}$).}
\label{fig:overview-labyrinth}
\end{figure}

This section gives an overview of our results on Labyrinth, a routing algorithm with shared state.
It introduces the algorithm, its sequential implementation
(\Cref{fig:overview-labyrinth-src}), and the dataflow graph our compiler produces from it
(\Cref{fig:overview-labyrinth-dfg}).

Labyrinth connects pairs of endpoints by \emph{routes} on a shared two-dimensional grid,
such that no two routes overlap. For each pair, the algorithm first \emph{searches} the 
grid for a candidate route, and then \emph{commits} the route by claiming its cells.
While the commit is cheap, the search is expensive and becomes a bottleneck at realistic scales.
For example, in chip layout, where the problem is known as (Lee-style) maze
routing~\cite{lee-routing}, benchmarks connect hundreds of thousands of
pairs~\cite{ispd18-routing} on grids of tens of millions of cells~\cite{ispd24-routing}.

Searching one route at a time is too slow at this scale, so Labyrinth runs the searches in
parallel, speculating that candidate routes rarely overlap. It works in rounds
(\Cref{fig:labyrinth-viz}). In each round, all searches run on the same snapshot of the grid.
Their candidate routes are then committed one after another, in the fixed order of the worklist.
A candidate route that crosses an already claimed cell is retried in the next round, on the
updated grid.

This shape of a round, speculative work on a snapshot followed by ordered commits and retries
on conflict, is common well beyond chip routing.
Databases and cloud services use it for optimistic concurrency control~\cite{optimistic-cc}:
transactions read a snapshot, validate their writes when they commit, and retry when validation fails,
as the candidate routes of Labyrinth do; a service that reserves limited stock for many concurrent
orders, for instance, works in the same way.
Similarly,
cluster schedulers place jobs on a snapshot of the cluster state and commit the placement
transactionally~\cite{omega}; parallel algorithms use it as deterministic
reservations~\cite{deterministic-reservations}, as do worklist-driven algorithms on shared data
structures in general~\cite{tao-of-parallelism}.
For these reasons, Labyrinth is part of the transactional-memory benchmark suite
STAMP~\cite{stamp}, and was used to evaluate \condrust~\cite{condrust}.

\paragraph*{Sequential implementation and reasoning.}
\Cref{fig:overview-labyrinth-src} shows the implementation of Labyrinth in our sequential
source language (\Cref{sec:source}).
Lines~\textcircled{\tiny 1} and~\textcircled{\tiny 2} allocate the two states of the algorithm:
the grid and the list of pairs to retry.
A \sletstate binds a \emph{state}: like a field of an object, or an owned mutable variable in
Rust, $\mathit{grid}$ is not passed around as a value but accessed only through calls such as
$\mathit{grid}.\mathsf{tryUpdate}(\mathit{cand})$.
Line~\textcircled{\tiny 3} starts the rounds: \strfix binds $\mathit{work}$ to the initial
worklist and runs its body once per round; the body sees only $\mathit{work}$ and the states in
scope.
A round (lines~\textcircled{\tiny 4}--\textcircled{\tiny 9}) takes a snapshot of the grid for the
pairs in $\mathit{work}$, and loops over the resulting search tasks in worklist order: it searches a
candidate route (\textcircled{\tiny 6}), tries to commit it (\textcircled{\tiny 7}), and records the
outcome (\textcircled{\tiny 8}).
Finally, $\mathit{retries}.\mathsf{nextWorklist}()$ ends the round
(\textcircled{\tiny 9}): it returns either $\mathsf{Continue}(\mathit{work}')$, with the pairs to
retry, which starts another round, or $\mathsf{Done}(v)$, which ends the recursion.
After the last round, $\mathit{grid}.\mathsf{routes}()$ returns the committed routes
(\textcircled{\tiny 10})\footnote{The implementations of Cao Minh et al.~\cite{stamp} and Suchert et al.~\cite{condrust} do not
return the routes. We return them so that the specification can state their disjointness: a postcondition
is a property of the value that the program returns.}.

\smallskip
We reason about the program of~\Cref{fig:overview-labyrinth-src} in a program logic for the source
language, which we built in Iris (\Cref{sec:logic}).
A specification of Labyrinth is a weakest precondition
$\WPsrc{\mathit{labyrinth}}{\Phi}.$ 
For $\Phi\triangleq\lambda v.\,\top$, the weakest precondition states \emph{safety}: Labyrinth never
reaches undefined behaviour, such as a \sfor over a value that is not a list, or a round that returns
neither $\mathsf{Done}$ nor $\mathsf{Continue}$.
For Labyrinth we prove more: with
\[
\Phi(v)\ \triangleq\ \forall r_1,r_2\in v.\ r_1\neq r_2\Rightarrow\mathsf{cells}(r_1)\cap\mathsf{cells}(r_2)=\emptyset,
\]
the committed routes are pairwise disjoint (\Cref{ex:labyrinth-verification}).
This is the key property in chip layout: two overlapping routes would be wires that short-circuit.
We prove it in Rocq twice, under the same specifications of the functions, which we also prove for a
concrete implementation: as partial correctness, and as total correctness, since every round commits or
drops at least one pair.

\subsection{The Dataflow Graph of Labyrinth}\label{sec:overview-compiler}

The program of \Cref{fig:overview-labyrinth-src} is verified, but it is sequential. It searches one
route after another, although the searches of a round are independent.
At realistic scales, Labyrinth must run in parallel, or distributed over many machines.
Even though Labyrinth can likely be implemented directly as a distributed program with the help of a 
large language model, (human or AI) reviewing or verifying its code
is challenging due to several factors: threads or cloud functions, locks or transactions
on the shared grid, and the retry protocol. 
Our compiler instead translates the sequential program into a dataflow graph
(\Cref{fig:overview-labyrinth-dfg}) whose execution is concurrent by construction, and ensures that the 
source specification -- e.g. absence of undefined behaviors or disjointness of committed routes above --
still holds for the distributed program.

\paragraph*{Construction.}
The compiler emits one node per call site and connects the nodes by \emph{channels}, the arrows of
\Cref{fig:overview-labyrinth-dfg}.
A channel is a FIFO queue from the node that writes it to the node that reads it; every
channel has exactly one writer and at most one reader.
A \emph{data channel} (solid arrows) carries a stream of items, ended by an end marker: for instance, the
channel from $\mathsf{For}$ to $\mathsf{findPath}$ carries the tasks of a round, one item per
iteration, and the channel from $\mathsf{findPath}$ to $\mathsf{tryUpdate}$ the candidate routes.
Inside a tail recursion, a stream is also divided into rounds by round markers.
A \emph{state channel} (dashed arrows) carries a single \emph{token}: the location of the state on the program
heap.
A node takes the token from its input state channel before its first call on the state, keeps it for all
the items of its input stream, or of the current round, and then passes it on to its output state channel.
The \sfor loop becomes a loop group: its $\mathsf{For}$ node turns the list of tasks into a stream, one item per
task, and every node of the loop body processes one item per iteration.
In particular, all searches of a round are firings of the single node $\mathsf{Stateless}(\mathsf{findPath})$.
Each state thus becomes a chain of state channels through the nodes that call it: the grid's token
passes from $\mathsf{StateInit}$, which allocates the grid, to $\mathsf{snapshot}$, to $\mathsf{tryUpdate}$,
and back, so a node can call the grid only while it holds the token.
The \strfix becomes a second loop group, headed by a $\mathsf{Fix}$ node.
$\mathsf{Fix}$ sends the current worklist into the graph of one round, and receives the round's result on
a feedback channel from $\mathsf{nextWorklist}$: on $\mathsf{Continue}(\mathit{work}')$ it starts the next
round with $\mathit{work}'$, on $\mathsf{Done}(v)$ it ends the recursion.
The tokens of $\mathit{grid}$ and $\mathit{retries}$ enter the round at its start and come back to
$\mathsf{Fix}$ at its end, so every round starts from the current states; after the last round they leave
$\mathsf{Fix}$, and the grid's token reaches $\mathsf{routes}$.

\paragraph*{Execution.}
The graph fixes only dependencies: a node can fire, that is, run one activation of its operator, as soon as the
items it needs are on its input channels.
The searches of a round need only their tasks, so they do not depend on each other; each
$\mathsf{tryUpdate}$ needs the grid's token, so the commits stay in worklist order.
The rounds run one after another, since each round starts from the result and the states of the previous
one; the parallelism is inside a round.
Running the searches of a round at the same time needs several copies of $\mathsf{findPath}$, which we
discuss in \Cref{sec:conclusions}.
The graph thus has exactly the dependencies of the sequential program: data flows along the data channels,
and the order of the calls on each state along its chain of state channels.

Which of the possible firings happens, and when, is decided by an \emph{executor}, and the same graph can
be run in two fundamentally different ways (\Cref{sec:target}):
\begin{itemize}
\item \emph{with a scheduler}, which fires one node at a time and so decides the order of all firings.
  The scheduler either follows a deterministic policy, for instance the order of the source program, or
  may choose any node that can make progress, which covers every interleaving of the firings;
\item \emph{without a scheduler}: any node whose inputs hold data may fire, and many nodes may fire at
  once, so the nodes can run in parallel, for instance on different machines, and no component knows the
  global order of the firings.
\end{itemize}

\subsection{Compilation Preserves the Specification of Labyrinth}\label{sec:overview-preservation}
The disjointness of the committed routes was proved for the sequential program; we do not prove it again
for the graph.
Our main theorem (\Cref{sec:verification}) states that compilation preserves weakest preconditions: from
$\WPsrc{\mathit{labyrinth}}{\Phi}$ it follows that the compiled graph satisfies $\Phi$ as well, for every
executor that meets an abstract specification of graph execution.
Intuitively, this specification \emph{passes a baton} between the nodes in the order of the source program:
calls and loop iterations need the baton, all other firings run freely, and the graph never deadlocks.
The execution scheduled to follow the order of the sequential program meets this specification, and so
does every other schedule, and the execution without a scheduler, with nodes firing in parallel.
These two results need weakest preconditions in which iterations are unguarded, that is, a proof of total
correctness, such as our second proof of Labyrinth.
Hence every run of the compiled graph of \Cref{fig:overview-labyrinth-dfg} that returns, in parallel and
without a scheduler, returns pairwise disjoint routes; we instantiate the theorems for Labyrinth in Rocq,
and running its compiled graph gives the same routes as the sequential program.

\section{Source Language}\label{sec:source}\label{sec:source-lang}\label{sec:fin-source}

\begin{figure}[t]
\[
\begin{array}{@{}r@{\ }c@{\ }l@{\qquad}l@{}}
v &::=& () \mid n \mid f \mid (v,v) \mid [v,\dots,v] & \text{values}\\[2pt]
e &::=& v \mid x \mid \mathbf{let}\ x=e\ \mathbf{in}\ e & \text{data}\\
  &\mid& f() \mid f(e) & \text{stateless calls}\\
  &\mid& s.f() \mid s.f(e) & \text{stateful calls}\\
  &\mid& \mathbf{letstate}\ s=e\ \mathbf{in}\ e & \text{state allocation}\\
  &\mid& \mathbf{for}\ x\ \mathbf{in}\ e\ \mathbf{do}\ e & \text{loop over a list}\\
  &\mid& \mathbf{trfix}\ x=e\ \mathbf{in}\ e & \text{tail recursion}\\[2pt]
r &::=& \mathsf{Done}(v) \mid \mathsf{Continue}(v) & \text{result of a \textbf{trfix} round}
\end{array}
\]
\caption{Source syntax. $x$ ranges over variables, $s$ over state names, $f$ over function names.
A stateful call names its state syntactically; every call takes at most one argument.}
\label{fig:source-syntax}
\end{figure}

Our source language is a small, first-order imperative language with state, loops and tail recursion
(\Cref{fig:source-syntax}).
Ordinary variables $x$ and state names $s$ live in separate namespaces.
A stateless call $f(e)$ depends only on its argument, while a stateful call $s.f(e)$ acts on the state
named $s$.
Control flow comes from the \sfor loop over a list and from the tail recursion \strfix.
A conditional is a \sfor over a list of zero or one elements, computed by a call: the body runs only
when the call returns a singleton.\footnote{Choices on values alone can be made inside the called
functions.}
In $\mathbf{trfix}\ x=e_i\ \mathbf{in}\ e_b$, the expression $e_i$ gives the first argument; each round runs
the body $e_b$ on the current argument and ends with $\mathsf{Done}(v)$, which stops the recursion with
result $v$, or with $\mathsf{Continue}(v')$, which starts another round with argument $v'$.

\subsection{ITree Semantics}\label{sec:source-sem}

\Cref{fig:source-sem} collects our \itree semantics for the source language.
The denotation of an expression $e$ is $\llbracket e\rrbracket_{\sigma,\rho}$, an \itree that returns a
value; the environment $\rho$ maps variables to values, and $\sigma$ maps state names to locations on the
program heap, which holds the values of the program's states, one cell per state.
A closed program starts with both $\rho$ and $\sigma$ empty.
The tree raises the events of $E_{\mathit{src}}$ (\Cref{fig:source-sem-events}): calls to the external
functions, operations on the program heap, undefined behaviour, and the start of a loop iteration.
What calls return, and what an iteration costs, is fixed only by the program logic (\Cref{sec:logic}).

The equations of \Cref{fig:source-sem-eqs} read as follows.
A stateful call $s.f(e)$ evaluates its argument and raises $\mathsf{sfcall}$ with the location
$\sigma(s)$ of its state; a stateless call is the same without the location.
\sletstate evaluates the initial value, allocates a fresh cell for it, and runs the body with $s$ bound
to the new location.
A \sfor evaluates the list and runs the body once per element, in order, with $x$ bound to the element;
each iteration starts with an $\mathsf{iter}$ event, and the loop returns $()$.
A \strfix evaluates the first argument and runs the body in the iteration combinator $\mathsf{loop}$
(\Cref{sec:background}): each round starts with an $\mathsf{iter}$ event, and its result
$\mathsf{Continue}(v')$ starts the next round with $v'$, while $\mathsf{Done}(v')$ ends the recursion
with $v'$.
The body sees only its argument ($\rho$ starts empty) but every state in scope ($\sigma$ is kept);
a \strfix that never returns $\mathsf{Done}$ is denoted by an infinite tree.
The $\mathsf{iter}$ events are where source and compiled graph meet loop by loop
(\Cref{sec:verification}).

\begin{figure}[t]
\small
\begin{subfigure}[t]{0.60\linewidth}
$\begin{array}[t]{@{}l@{\ }c@{\ }l@{}}
\multicolumn{3}{@{}l@{}}{\llbracket e\rrbracket_{\sigma,\rho}\ :\ \itreeER{E_{\mathit{src}}}{\mathsf{val}}}\\[2pt]
\multicolumn{3}{@{}l@{}}{\llbracket e\rrbracket\ \triangleq\ \llbracket e\rrbracket_{\emptyset,\emptyset}}\\[6pt]
\llbracket s.f(e)\rrbracket_{\sigma,\rho} &=& v\leftarrow\llbracket e\rrbracket_{\sigma,\rho};\
  \mathsf{sfcall}(f,\sigma(s),[v])\\[6pt]
\llbracket\mathbf{letstate}\ s=e_i\ \mathbf{in}\ e_b\rrbracket_{\sigma,\rho} &=& v\leftarrow\llbracket e_i\rrbracket_{\sigma,\rho};\\
  && \ell\leftarrow\mathsf{alloc}(v);\ \llbracket e_b\rrbracket_{\sigma[s\mapsto\ell],\rho}\\[6pt]
\llbracket\mathbf{for}\ x\ \mathbf{in}\ e_i\ \mathbf{do}\ e_b\rrbracket_{\sigma,\rho} &=& [v_1,\dots,v_n]\leftarrow\llbracket e_i\rrbracket_{\sigma,\rho};\\
  && \mathsf{iter};\ \llbracket e_b\rrbracket_{\sigma,\rho[x\mapsto v_1]};\ \dots\\
  && \mathsf{iter};\ \llbracket e_b\rrbracket_{\sigma,\rho[x\mapsto v_n]};\ \mathsf{Ret}\ ()\\[6pt]
\llbracket\mathbf{trfix}\ x=e_i\ \mathbf{in}\ e_b\rrbracket_{\sigma,\rho} &=& v_0\leftarrow\llbracket e_i\rrbracket_{\sigma,\rho};\\
  && \mathsf{loop}\ \bigl(\lambda v.\ \mathsf{iter};\ \llbracket e_b\rrbracket_{\sigma,[x\mapsto v]}\bigr)\ v_0
\end{array}$
\caption{\itree denotation, selected equations (all of them in \Cref{app:src-sem}).}\label{fig:source-sem-eqs}
\end{subfigure}\hfill{\color{black!20}\vrule width 0.4pt}\hfill
\begin{subfigure}[t]{0.36\linewidth}
$\begin{array}[t]{@{}l@{}}
\makebox[0pt][l]{$E_{\mathit{src}}\ \triangleq\ \mathsf{funCallE}+\mathsf{progE}+\mathsf{ubE}+\mathsf{iterE}$}\\[4pt]
\mathsf{slcall}(f,\bar v)\ :\ \mathsf{funCallE}\ \mathsf{val}\\[4pt]
\mathsf{sfcall}(f,\ell,\bar v)\ :\ \mathsf{funCallE}\ \mathsf{val}
\end{array}$
\caption{Event signature.}\label{fig:source-sem-events}
\end{subfigure}
\caption{\itree semantics of the source language.
$\mathsf{funCallE}$ contains the calls to the external functions: a stateless call raises
$\mathsf{slcall}(f,\bar v)$, a stateful call raises $\mathsf{sfcall}(f,\ell,\bar v)$ with the location
$\ell$ of its state, and both are answered with the result of the call.
$\mathsf{progE}$ contains the standard operations on the program heap, $\mathsf{alloc}$, $\mathsf{load}$
and $\mathsf{store}$; the source raises only $\mathsf{alloc}(v)$, answered with a fresh location.
$\mathsf{ubE}$ contains the single event $\mathsf{ub}$, which signals undefined behaviour, such as an
unbound variable, a \sfor over a value that is not a list, or a \strfix round that returns neither
$\mathsf{Done}$ nor $\mathsf{Continue}$; it has no answer.
$\mathsf{iterE}$ contains the single event $\mathsf{iter}$, which marks the start of a loop iteration.}
\label{fig:source-sem}
\end{figure}

\subsection{Program Logic}\label{sec:logic}
Following Vistrup et al.~\cite{plalacarte}, we obtain a program logic for the source language from its
\itree semantics by giving each signature of $E_{\mathit{src}}$ a logical handler
(\Cref{sec:background}).
We write $H(\epsilon,\Psi)$ for what a handler requires before the event $\epsilon$, so that the program
continues with the postcondition $\Psi$ on the answer.
The functions that a program calls are known only through their specifications.
A specification $\mathsf{spec}_f$ relates arguments and results, and, for a stateful function, the old
and the new value of the state: $\mathsf{spec}_f(w,\bar v,w',r)$ says that $f$, called with arguments
$\bar v$ on a state with value $w$, may leave the state with value $w'$ and return $r$.
The handler of calls lets the program continue with every answer that the specification allows:
\[
\begin{array}{@{}r@{\ }c@{\ }l@{}}
H_{\mathit{fun}}(\mathsf{slcall}(f,\bar v),\Psi) &\triangleq&
  \forall r.\ \mathsf{spec}_{f}(\bar v,r)\mathrel{-\!\!*}\Psi(r),\\[2pt]
H_{\mathit{fun}}(\mathsf{sfcall}(f,\ell,\bar v),\Psi) &\triangleq&
  \exists w.\ \ell\mapsto w\ast\bigl(\forall w',r.\ \ell\mapsto w'\mathrel{-\!\!*}
  \mathsf{spec}_{f}(w,\bar v,w',r)\mathrel{-\!\!*}\Psi(r)\bigr).
\end{array}
\]
So a proof holds for every implementation of $f$ that meets its specification.
A stateful call needs the points-to assertion $\ell\mapsto w$ of its state, which gives exclusive
ownership of the cell at $\ell$, holding $w$; it reads the old value from it and gives it back with the
new value that the specification allows.
The handler of the program heap gives the usual rules of separation logic (\Cref{app:src-handlers}); for the only event that the
source raises, $H_{\mathit{heap}}(\mathsf{alloc}(v),\Psi)\triangleq\forall\ell.\ \ell\mapsto
v\mathrel{-\!\!*}\Psi(\ell)$, so \sletstate produces the points-to of a fresh location.
A proof of a program therefore tracks the value of every state, as the invariant of
\Cref{ex:labyrinth-verification} does.
Triggering undefined behaviour requires a proof of false,
$H_{\mathit{ub}}(\mathsf{ub},\Psi)\triangleq\bot$~\cite{plalacarte}.
The handler of iterations is left open, and the user of the logic chooses it.
With $H_{\mathit{iter}}(\mathsf{iter},\Psi)\triangleq\triangleright\Psi(())$, every iteration guards the
rest of the program by a later, and L\"ob induction proves properties of a \strfix that need not
terminate: partial correctness.
With $H_{\mathit{iter}}(\mathsf{iter},\Psi)\triangleq\Psi(())$, iterations are \emph{unguarded}; since weakest
preconditions are least fixpoints, a proof of a \strfix then needs a measure that decreases at every
round: total correctness.
Our preservation result holds for every choice, while its extensions to every schedule and to execution
without a scheduler need unguarded iterations (\Cref{sec:verification}).
The handler of the source event signature $E_{\mathit{src}}$ is therefore
$
H_{\mathit{src}}\triangleq H_{\mathit{fun}}\oplus H_{\mathit{heap}}\oplus H_{\mathit{ub}}\oplus
H_{\mathit{iter}},
$
and the weakest precondition of an expression is the one of its denotation,
$\WPsrc{e}{\Phi}\triangleq\wpi{H_{\mathit{src}}}{\llbracket e\rrbracket}{\Phi}$, required for every
instance of the ghost state of the program heap (\Cref{app:src-wp}); we write $\WPsrctot{e}{\Phi}$ when
iterations are unguarded.
Since $H_{\mathit{ub}}$ requires $\bot$, $\WPsrc{e}{\lambda v.\,\top}$ already states that $e$ is
\emph{safe}: it never reaches undefined behaviour.

\begin{example}[Partial and total correctness for Labyrinth]\label[example]{ex:labyrinth-verification}
We prove that the program of \Cref{fig:overview-labyrinth-src} commits pairwise disjoint routes,
$\vdash\WPsrc{\mathit{labyrinth}}{\Phi}$ with
\[
\Phi(v)\ \triangleq\ \forall r_1,r_2\in v.\ r_1\neq r_2\Rightarrow\mathsf{cells}(r_1)\cap\mathsf{cells}(r_2)=\emptyset.
\]
The functions stay abstract. Writing $\mathsf{routes}(g)$ for the routes committed to a grid $g$ and
$\mathsf{claimed}(g)$ for the union of their cells, and $c\mathrel{\#}g$ when the cells of the route $c$
are all unclaimed in $g$, $\mathsf{cells}(c)\cap\mathsf{claimed}(g)=\emptyset$, we prove it under these
specifications:\footnote{These specifications are not assumed: we give a concrete implementation of every
function and prove that it meets them.}
\[
\begin{array}{@{}l@{\ }l@{}}
\mathsf{emptyGrid}: & \text{returns } g \text{ with } \mathsf{routes}(g)=\emptyset\\[2pt]
\mathsf{emptyList}: & \text{returns } [\,]\\[2pt]
\mathsf{initialWorklist}: & \text{returns a list}\\[2pt]
\mathsf{snapshot}: & g\xrightarrow{[p_1,\dots,p_n]}g,\ [(p_1,g),\dots,(p_n,g)]\\[2pt]
\mathsf{findPath}: & (p,g)\mapsto\mathsf{NoPath}(p)\ \text{ or }\ \mathsf{Cand}(p,c)\ \text{ with }\ c\mathrel{\#}g\\[2pt]
\mathsf{tryUpdate}: & g\xrightarrow{\mathsf{NoPath}(p)}g,\ \mathsf{Dropped}\\
  & g\xrightarrow{\mathsf{Cand}(p,c)}g',\ \mathsf{Committed}\ \text{ with }\
    \mathsf{routes}(g')=\mathsf{routes}(g)\cup\{c\},\ \text{ if }\ c\mathrel{\#}g\\
  & g\xrightarrow{\mathsf{Cand}(p,c)}g,\ \mathsf{Retry}(p)\ \text{ otherwise}\\[2pt]
\mathsf{record}: & \mathit{rs}\xrightarrow{o}\mathit{rs}\mathbin{+\!\!+}[p]\ \text{ if } o=\mathsf{Retry}(p),
  \text{ and } \mathit{rs} \text{ otherwise}\\[2pt]
\mathsf{nextWorklist}: & \mathit{rs}\xrightarrow{}[\,],\ \mathsf{Done}(v)\ \text{ if } \mathit{rs}=[\,],
  \text{ and } \mathsf{Continue}(\mathit{rs}) \text{ otherwise}\\[2pt]
\mathsf{routes}: & g\xrightarrow{}g,\ \mathsf{routes}(g)
\end{array}
\]
where $g\xrightarrow{a}g',r$ reads ``called with argument $a$ on state $g$, the function may leave the
state $g'$ and return $r$''.

The proof is an ordinary sequential proof, one call at a time. At the start of every round it keeps the
invariant
\[
I\ \triangleq\ \exists g.\ \ell_{\mathit{grid}}\mapsto g\ast\ulcorner\forall r_1,r_2\in\mathsf{routes}(g).\
  r_1\neq r_2\Rightarrow\mathsf{cells}(r_1)\cap\mathsf{cells}(r_2)=\emptyset\urcorner
  \ast\ell_{\mathit{retries}}\mapsto[\,],
\]
where $\ell_{\mathit{grid}}$ and $\ell_{\mathit{retries}}$ are the locations that \sletstate allocates for
$\mathit{grid}$ and $\mathit{retries}$.
$I$ holds after line~\textcircled{\tiny 1}, since the initial grid has no routes and the initial list of
retries is empty.
Within a round, $\mathsf{snapshot}$ leaves the grid unchanged, $\mathsf{tryUpdate}$ either adds a route
whose cells are all unclaimed, which keeps the routes disjoint, or leaves the grid unchanged, and
$\mathsf{record}$ collects the retried pairs; $\mathsf{nextWorklist}$ then empties $\mathit{retries}$ and
passes its contents to the next round.
So every round goes from $I$ to $I$.
Moreover, the first candidate of a round was searched on the grid that $\mathsf{tryUpdate}$ sees, since
nothing is committed before it, so it commits or is dropped: a round on $n>0$ pairs retries at most $n-1$
of them, and a round on no pair returns $\mathsf{Done}$.

\smallskip
\emph{Partial correctness.}
With a later per iteration, the recursive occurrence of the round is guarded, and L\"ob induction proves
the \strfix under the hypothesis that all later rounds preserve $I$; the progress of a round is not needed.
Finally, $\mathsf{routes}$ returns $\mathsf{routes}(g)$ for the grid $g$ of $I$, which establishes $\Phi$:
if Labyrinth terminates, its routes are disjoint.
This proof gives a weakest precondition with guarded iterations, for which we have no adequacy theorem;
the facts about runs below rest on the second proof.

\smallskip
\emph{Total correctness.}
With unguarded iterations, a proof of a \strfix needs a measure that decreases at every round.
The length of the worklist is one, by the progress of a round, and the same argument proves
$\WPsrctot{\mathit{labyrinth}}{\Phi}$: Labyrinth terminates with disjoint routes.
Since the specifications are proved for a concrete implementation, the resulting statement has no
hypotheses: the program runs without undefined behaviour and returns disjoint routes, and, by the results
of \Cref{sec:verification}, so does its compiled graph (\Cref{sec:overview-preservation}).
\end{example}

\section{Target Language}\label{sec:target}

The target of our compiler is a dataflow language. A program of this language first
\emph{constructs} a graph, allocating its channels and creating its nodes, and then
\emph{executes} it. Construction has an \itree semantics, in which the execution of a graph is a single event; an
\emph{executor} gives that event its meaning, by interpreting the program's \itree and replacing each
execution event with a run of the graph. With this approach the same graph can be run by different
executors (\Cref{sec:target-exec}).

\subsection{Syntax}\label{sec:target-syntax}

\begin{figure}[t]
\[
\begin{array}{@{}r@{\ }c@{\ }l@{\qquad}l@{}}
g &::=& x \mid v \mid \mathbf{chan} \mid \mathbf{fst}\ g \mid \mathbf{snd}\ g
       \mid \mathbf{let}\ x=g\ \mathbf{in}\ g & \text{construction}\\
  &\mid& k(\bar g) & \text{node of kind } k\\
  &\mid& \mathbf{loop}(\bar g) & \text{loop group}\\
  &\mid& \mathbf{execute}(\bar g) & \text{execution}\\[4pt]
k &::=& \mathsf{Const} \mid \mathsf{Value} \mid \mathsf{Source}(f) \mid \mathsf{Fanout} \mid \mathsf{Reuse}
       & \text{data nodes}\\
  &\mid& \mathsf{For} \mid \mathsf{Fix}(n_p) & \text{loop heads}\\
  &\mid& \mathsf{Stateless}(f) \mid \mathsf{StateInit} \mid \mathsf{Stateful}_0(f) \mid \mathsf{Stateful}(f)
       & \text{call and state nodes}\\
  &\mid& \mathsf{Target} & \text{result}\\[4pt]
o &::=& x \mid v \mid \mathbf{let}\ x=o\ \mathbf{in}\ o \mid \mathbf{if}\ o\ \mathbf{then}\ o\ \mathbf{else}\ o
       \mid \dots & \text{operators}\\
  &\mid& \mathbf{recv}(c) \mid \mathbf{send}(c,o) & \text{communication}\\
  &\mid& f(\bar o) \mid o.f(\bar o) \mid \mathbf{newstate}(o) & \text{program effects}\\
  &\mid& \mathbf{ref}\ o \mid\ !o \mid o:=o & \text{local references}
\end{array}
\]
\caption{Syntax of the dataflow graph language. $\mathbf{chan}$ allocates a channel, a first-in first-out
queue, and returns the pair of its sending and receiving end, which $\mathbf{fst}$ and $\mathbf{snd}$
select; $\mathbf{let}$ binds channels and nodes. $k(\bar g)$ creates a node of kind $k$ on the channel
ends it reads and writes (and, for $\mathsf{Const}$ and $\mathsf{Value}$, a constant).
$\mathbf{loop}(\bar g)$ groups the nodes of one loop. $\mathbf{execute}(\bar g)$ runs the nodes $\bar g$
and returns the result of the graph. The last block is the operator language, in which the operator of
every node kind is written: a small functional core, communication on channels, function calls and state
allocation, and local references.}
\label{fig:target-syntax}
\end{figure}

\begin{figure}[t]
\centering
\begin{subfigure}[c]{0.46\textwidth}
\small
\[
\begin{array}{@{}l@{}}
\mathbf{let}\ l,i,x,k,z,r,t_0,t_1,t_2=\mathbf{chan}\ \mathbf{in}\\
\mathbf{execute}(\mathsf{Const}(0)(\to i),\ \mathsf{StateInit}(i\to t_0),\\
\quad\mathsf{Const}([1,2])(\to l),\\
\quad\mathbf{loop}(\mathsf{For}(l\to x,k),\\
\quad\phantom{\mathbf{loop}(}\mathsf{Stateful}(\mathsf{add})(t_0,x\to z,t_1)),\\
\quad\mathsf{Stateful}_0(\mathsf{get})(t_1,k\to r,t_2),\ \mathsf{Target}(r))
\end{array}
\]
\caption{Program implemented in the syntax of the graph language.}\label{fig:target-example-prog}
\end{subfigure}\hfill
\begin{subfigure}[c]{0.50\textwidth}
\centering
\begin{tikzpicture}[x=1cm,y=0.9cm]
  \begin{scope}[on background layer]
    \draw[rounded corners=3pt,fill=fLoop,draw=black!40] (1.75,-0.95) rectangle (4.55,-3.65);
  \end{scope}
  \node[flab,anchor=south east] at (4.55,-3.65) {$\mathbf{loop}$};
  \node[fnode] (c0) at (0,0) {$\mathsf{Const}(0)$};
  \node[fnode] (si) at (0,-1.6) {$\mathsf{StateInit}$};
  \node[fnode] (cl) at (3,0) {$\mathsf{Const}([1,2])$};
  \node[fnode] (fo) at (3,-1.6) {$\mathsf{For}$};
  \node[fnode] (ad) at (3,-2.9) {$\mathsf{Stateful}(\mathsf{add})$};
  \node[fnode] (gt) at (1.4,-4.7) {$\mathsf{Stateful}_0(\mathsf{get})$};
  \node[fnode] (tg) at (1.4,-6.0) {$\mathsf{Target}$};
  \draw[fdata] (c0) -- node[flab,right]{$i$: $0$} (si);
  \draw[fdata] (cl) -- node[flab,right]{$l$: $[1,2]$} (fo);
  \draw[fdata] (fo) -- node[flab,right]{$x$: $1,2$} (ad);
  \draw[fdata] (fo.east) -- (5.6,-1.6) |- node[flab,right,pos=0.3]{$k$: $2$} (gt.east);
  \draw[fdang] (ad.east) -- (4.95,-2.9) node[flab,right]{$z$};
  \draw[fdata] (gt) -- node[flab,right]{$r$} (tg);
  \draw[fstate] (si.south) |- node[flab,above,pos=0.75]{$t_0$} (ad.west);
  \draw[fstate] (ad.south) -- ++(0,-0.75) -| node[flab,above,pos=0.25]{$t_1$} (gt.north);
  \draw[fstate,thin,draw=fState!50] (gt.west) -- ++(-0.7,0) node[flab,left]{$t_2$};
\end{tikzpicture}
\caption{Visual representation of the program on the left.}\label{fig:target-example-graph}
\end{subfigure}
\caption{A graph program that adds the elements of $[1,2]$ to a
state that starts at $0$ and returns its final value (a), and its visual representation (b). In (a), one $\mathbf{let}$ binds
several channels, a channel's name stands for both of its ends, and every node lists its inputs before the arrow and its outputs after it.
In (b), as in \Cref{fig:overview-labyrinth}: solid, data; dashed, state; thin, outputs that no node reads ($z$, the results of $\mathsf{add}$, and $t_2$, the last token of the state).
Labels give the items a channel carries before its end marker.}
\label{fig:target-example}
\end{figure}
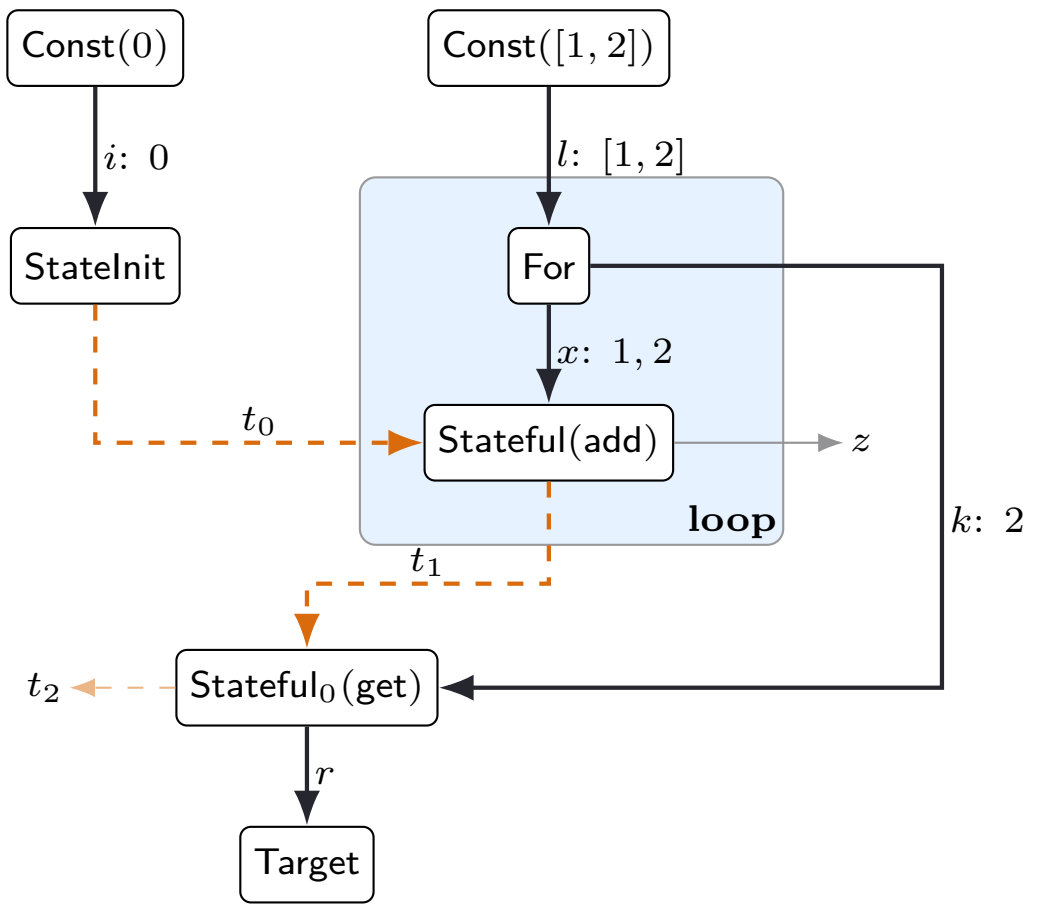

A graph program allocates channels, creates nodes on them, and then executes the nodes
(\Cref{fig:target-syntax}).
The graph program of \Cref{fig:target-example} adds the elements of the
list $[1,2]$ to a state that starts at $0$, and returns the final value of the state. It allocates nine
channels and then executes seven nodes, each applied to the channels it reads and writes;
\Cref{fig:target-example-graph} draws the graph.

Channels are untyped: what a channel carries depends only on the nodes at its two ends.
Most channels carry a \emph{stream}: a sequence of items, ended by an end marker (\Cref{app:streams}); we call them
\emph{data channels}. When the items of a stream only say when a node acts, we call the stream a
\emph{trigger}.
A \emph{state channel} instead carries the \emph{token} of a state, its location on the program heap
(\Cref{sec:overview-compiler}), from one node that calls the state to the next; inside a $\mathsf{Fix}$
loop (below) it carries one token per round.
In the example, $l$ carries one item, the list $[1,2]$, and then the end marker, and $t_0$, $t_1$, $t_2$
are the state channels of the state.

Every node kind is implemented by an \emph{operator}: a small sequential program in the operator language
of \Cref{fig:target-syntax}, fixed by the language, which the programmer never writes
(\Cref{app:op-lang,app:operators} give the language and all operators).
A node runs its operator as the body of a loop (\Cref{sec:background}), and each run is an
\emph{activation}. An activation reads one item from every input it needs, may call one function, sends
on the outputs, and returns $\mathsf{Continue}(s)$, with the \emph{node state} $s$ for the next
activation, or $\mathsf{Done}$, when the node has finished.
For instance, the operator of $\mathsf{Stateful}(\mathsf{add})$ in \Cref{fig:target-example}, which
reads the token on $t_0$ and the items on $x$, and writes on $z$ and $t_1$, is
\[
\begin{array}{@{}l@{}}
\mathbf{let}\ \ell=\mathbf{if}\ s=\bot\ \mathbf{then}\ \mathbf{recv}(t_0)\ \mathbf{else}\ s\ \mathbf{in}\\
\mathbf{let}\ w=\mathbf{recv}(x)\ \mathbf{in}\\
\mathbf{if}\ \mathsf{isEnd}(w)\ \mathbf{then}\ \mathbf{send}(z,\mathsf{end});\ \mathbf{send}(t_1,\ell);\ \mathsf{Done}\\
\mathbf{else}\ \mathbf{send}(z,\ell.\mathsf{add}(w));\ \mathsf{Continue}(\ell)
\end{array}
\]
Its node state is the token $\ell$, initially absent ($\bot$). The first activation takes the token from
$t_0$ and calls $\mathsf{add}(1)$, the second keeps the token and calls $\mathsf{add}(2)$, and the third
reads the end marker and passes the token on $t_1$.

Besides the loop heads, which we describe with loops below, the node kinds fall into three groups.
Data nodes move items: $\mathsf{Const}$ sends one constant, $\mathsf{Value}$ sends a constant for every
item of a trigger, and $\mathsf{Fanout}$ copies a stream to several channels; $\mathsf{Reuse}$ belongs to
loops.
Call and state nodes make function calls and allocate states: $\mathsf{Source}(f)$ calls $f()$ for every item of a
trigger, and $\mathsf{Stateless}(f)$ calls $f$ on every item of its input. $\mathsf{StateInit}$ reads one
item, allocates a state with it on the program heap, and sends the state's token.
$\mathsf{Stateful}_0(f)$ and $\mathsf{Stateful}(f)$ receive the token of a state, call $f$ on the state
for every item of a trigger and on every item of their input, respectively, and then pass the token on:
the state channels of one state form a chain that orders its calls.
In the example, $\mathsf{Stateful}_0(\mathsf{get})$ can call $\mathsf{get}$ only after
$\mathsf{Stateful}(\mathsf{add})$ has passed the token on $t_1$, that is, after both calls to
$\mathsf{add}$.
Finally, $\mathsf{Target}$ returns the single item of its input as the result of the graph.

A \emph{loop group} collects the nodes of one loop, and its head is a $\mathsf{For}$ or a
$\mathsf{Fix}$ node. The group adds no computation, but makes the loop visible to the
executor (\Cref{sec:target-exec}).
$\mathsf{For}$ runs one iteration per element of a list: for every list on its input, it sends the elements of the
list, one item each, and on a second channel the length of the list, its \emph{extent}. In the
example, $\mathsf{For}$ sends $1$ and $2$ on $x$, and the extent $2$ on $k$, which serves as the trigger
of $\mathsf{get}$.
$\mathsf{Reuse}$ repeats every item of a stream as often as an extent says, so that a value computed
outside the loop reaches every iteration.

$\mathsf{Fix}(n_p)$ heads a loop that runs in rounds and passes the tokens of $n_p$ states from each round
to the next; the body of the loop is an ordinary subgraph in the same loop group.
Inside the body, streams are split into \emph{rounds}, one per iteration, by \emph{round markers}. A node
treats a round marker as the end of its stream, except that it passes on a round marker and then starts
over for the next round; the operator of $\mathsf{Stateful}(\mathsf{add})$ above has a third case for it, which we
omitted. Outside the body of a $\mathsf{Fix}$ loop, no round marker occurs.
$\mathsf{Fix}$ first receives the tokens of its states, and then reads its arguments, the
\emph{seeds}, one at a time.
A seed $a$ starts a round: $\mathsf{Fix}$ sends the tokens into the body, and $a$ followed by a round
marker on the body's argument channel. At the end of the round, it receives the result of the round on a
feedback channel and the tokens back from the body.
On $\mathsf{Continue}(a')$ it starts the next round from $a'$; on $\mathsf{Done}(v)$ it sends $v$ on its
result channel and reads the next seed. At the end marker of its seeds, it sends the tokens on to the rest
of the graph.
When the $\mathsf{Fix}$ loop is itself inside the body of a loop, its seeds are split into rounds too: on a round marker,
$\mathsf{Fix}$ ends the outer round, sends the tokens on, and receives the tokens of the next round.
The feedback channel and the channels that bring the tokens back form cycles in the graph.

\subsection{Semantics}\label{sec:target-sem}

A graph program $g$ is denoted by an \itree $\llbracket g\rrbracket:\itreeER{E_{\mathit{dfg}}}{\mathsf{val}}$,
whose events describe how the graph is built and run:
\[
\begin{array}{@{}r@{\ }c@{\ }l@{}}
E_{\mathit{dfg}} &\triangleq& \mathsf{constrE}+\mathsf{execE}+\mathsf{ubE}+\mathsf{stepE}\\[4pt]
\mathsf{chan} &:& \mathsf{constrE}\ (\mathsf{chan}\times\mathsf{chan})\\
\mathsf{node}(n) &:& \mathsf{constrE}\ \mathsf{val}\\
\mathsf{execute}(\bar n) &:& \mathsf{execE}\ \mathsf{val}
\end{array}
\]
The \emph{construction} events $\mathsf{constrE}$ build the graph: $\mathsf{chan}$ asks for a fresh channel
and is answered with its two ends, and $\mathsf{node}(n)$ creates the node $n$, a node kind applied to its
channels, or a loop group of nodes.
The \emph{execution} event $\mathsf{execute}(\bar n)$ runs the graph formed by the nodes $\bar n$ and is
answered with the result.
As in the source, $\mathsf{ub}$ signals undefined behaviour, for instance a node applied to something
that is not a channel; $\mathsf{step}$ marks a step of computation without an observable effect, which
executors raise (\Cref{sec:target-exec}).
The equations are structural (\Cref{app:tgt-sem}): $\mathbf{chan}$ raises $\mathsf{chan}$, a node or a loop group raises
$\mathsf{node}$ once its arguments are evaluated, and $\mathbf{execute}(\bar g)$ evaluates the nodes and
raises $\mathsf{execute}(\bar n)$.
The execution of a whole graph is thus a single event, and its meaning is given by an executor
(\Cref{sec:target-exec}).

An executor runs a graph activation by activation, and every activation runs the operator of its node,
which is denoted by an \itree over the events
$E_{\mathit{op}}\triangleq\mathsf{commE}+\mathsf{funCallE}+\mathsf{progE}+\mathsf{ubE}+\mathsf{stepE}$.
The \emph{communication} events $\mathsf{commE}$ are the operations on channels:
$\mathsf{send}(c,v):\mathsf{commE}\ \mathsf{unit}$ appends $v$ to the queue of $c$, and $\mathsf{recv}(c):\mathsf{commE}\ \mathsf{val}$ removes the first item of the queue and is
answered with it. A $\mathsf{recv}$ on an empty channel is undefined behaviour; executors never cause one,
because they run an activation only when the inputs it reads hold data (\Cref{sec:target-exec}).
$\mathsf{funCallE}$, $\mathsf{progE}$ and $\mathsf{ubE}$ are the events of the source
(\Cref{sec:source-sem}); the source's $\mathsf{iter}$ is raised by the executor, not by operators, and
$\mathsf{step}$ is as in $E_{\mathit{dfg}}$. This is where a graph meets its source program: an operator makes the same function
calls as the source, allocates states on the same program heap, and raises the same $\mathsf{ub}$.

The graph thus works on three heaps: the program heap of the states, shared with the source; the
\emph{communication heap} of the channel queues; and the \emph{operator heap} of the local references,
accessed with $\mathbf{ref}$, $!$, and $:=$.
Only the first one is visible outside the graph. Local references raise no event outside the operator, and
no node needs to share them with another: the node state passes from one activation to the next as a
value.

\subsection{Execution}\label{sec:target-exec}

An executor gives the execution event its meaning: it interprets the \itree of a graph program and
replaces every event $\mathsf{execute}(\bar n)$ with a run of the nodes $\bar n$.
In a run, nodes \emph{fire}, and every firing is one activation of the node's operator
(\Cref{sec:target-syntax}), as for the actors of dataflow process networks~\cite{lee-parks-dpn}.
All our executors share three rules. A node fires only when the inputs that its next activation reads
hold data, so no $\mathsf{recv}$ finds an empty channel (\Cref{sec:target-sem}).
The executor raises $\mathsf{iter}$ after every activation that starts an iteration: one of a
$\mathsf{For}$ that sends an element, and one of a $\mathsf{Fix}$ that starts a round, by taking a seed
or by receiving $\mathsf{Continue}$; receiving $\mathsf{Done}$ starts no iteration.
Finally, the executor raises $\mathsf{step}$ events, which mark its own progress.
If no node can fire before the graph has returned its result, the run is stuck, and the executor raises
$\mathsf{ub}$.
\Cref{app:executors} defines the executors below precisely.
The executors differ only in who decides which node fires next, and in how many nodes fire at once.

\emph{With a scheduler.} A scheduler fires one node per round of the executor, and every round raises one $\mathsf{step}$.
A \emph{policy} decides which of the nodes that can fire fires next.
The \emph{source-order} policy fires the next function call or iteration of the source program as soon
as its node can fire, and some other node otherwise; under it, a compiled graph makes its calls and
iterations in the order of the source.
The \emph{demonic} scheduler instead lets the environment choose any node that can fire, so its runs are
all the schedules of the graph.
Both keep, beside the channels, the position of the run in the source order: the policy reads it, and
the demonic scheduler keeps it only for the proofs (\Cref{sec:verification}).

\emph{Without a scheduler.} In a distributed execution, no component decides, or even knows, the global order of the firings: a
node fires as soon as its inputs hold data. We model this in two ways.
The \emph{local} executor keeps no state beside the channels: in every round, any node that can fire may
fire, and the choice is demonic. It is the demonic scheduler with its bookkeeping erased, so the two have
the same runs, up to $\mathsf{Tau}$ steps.
The \emph{parallel} executor fires many nodes at once. In every round, it takes a snapshot of the
channels, computes a \emph{batch} of nodes that can fire and read pairwise disjoint inputs, greedily in
the order of the graph, and fires them all on the snapshot; then, on every channel, it removes the items
the batch received and appends the items it sent, in batch order.
Since the batch depends only on the channels, the parallel executor makes no scheduling choice of its
own; the arbitrary timings of a distributed execution are those of the local executor.
It raises one $\mathsf{step}$ per activation, so a batch of $k$ nodes has the events of $k$ rounds of the
local executor.

\begin{example}\label[example]{ex:target-exec}
\Cref{tab:target-exec} runs the graph of \Cref{fig:target-example} under each executor.
Its sixteen activations are the same in every run, and so are the calls: $\mathsf{add}(1)$,
$\mathsf{add}(2)$, and $\mathsf{get}$, in this order, because $x$ delivers the elements in order and the
token of the state passes from $\mathsf{add}$ to $\mathsf{get}$.
The iterations, instead, move: under the source-order policy, every iteration comes right before its
call, as in the source, while the other executors may let $\mathsf{For}$ run ahead of
$\mathsf{Stateful}(\mathsf{add})$.
The parallel executor needs nine rounds, since independent nodes, such as the last activation of
$\mathsf{For}$ and the first of $\mathsf{Stateful}(\mathsf{add})$, fire in the same batch.
\end{example}

\begin{table}[t]
\centering
\small
\begin{tabular}{@{}l>{\raggedright\arraybackslash}p{0.5\textwidth}l@{}}
\hline
Executor & A run & Calls and iterations\\
\hline
source order & $C_1\,C_2\,I_1\,I_2\,L_1\,F_1\,A_1\,F_2\,A_2\,L_2\,F_3\,A_3\,G_1\,G_2\,T_1\,T_2$
  & $\mathsf{iter}\,\mathsf{add}(1)\,\mathsf{iter}\,\mathsf{add}(2)\,\mathsf{get}$\\
demonic, local & $C_1\,C_2\,I_1\,I_2\,L_1\,L_2\,F_1\,F_2\,F_3\,A_1\,A_2\,A_3\,G_1\,G_2\,T_1\,T_2$
  & $\mathsf{iter}\,\mathsf{iter}\,\mathsf{add}(1)\,\mathsf{add}(2)\,\mathsf{get}$\\
parallel & $\{C_1,L_1\}$ $\{C_2,I_1,L_2,F_1\}$ $\{I_2,F_2\}$ $\{F_3,A_1\}$ $\{A_2\}$ $\{A_3\}$ $\{G_1\}$ $\{G_2,T_1\}$ $\{T_2\}$
  & $\mathsf{iter}\,\mathsf{iter}\,\mathsf{add}(1)\,\mathsf{add}(2)\,\mathsf{get}$\\
\hline
\end{tabular}
\caption{Runs of the graph of \Cref{fig:target-example} under each executor of \Cref{sec:target-exec}. The
graph allocates a state that starts at $0$, adds to it the elements of the list $[1,2]$, and then reads it.
Its nodes are $C=\mathsf{Const}(0)$, $L=\mathsf{Const}([1,2])$, $I=\mathsf{StateInit}$, $F=\mathsf{For}$,
$A=\mathsf{Stateful}(\mathsf{add})$, $G=\mathsf{Stateful}_0(\mathsf{get})$ and $T=\mathsf{Target}$, and
$n_j$ is the $j$-th activation of node $n$. A node fires once for every item it handles and once more for
the end marker: $C_1$ sends $0$ and $C_2$ the end marker; $F_1$ takes the list and sends $1$, $F_2$ sends
$2$, and $F_3$ reads the end marker. The demonic and the local executor allow every order in which a node
fires only when its inputs hold data, and the source-order policy fixes the order of the calls and
iterations but chooses the place of the other activations; we show one run each. A parallel run is a
sequence of batches $\{\dots\}$, whose nodes fire together. The last column lists, in order, the calls and
the events $\mathsf{iter}$, which $\mathsf{For}$ raises when it sends an element.}
\label{tab:target-exec}
\end{table}

\section{Compiler}\label{sec:compiler}

This section presents our compiler and its proof.
The compiler translates a program of the source language (\Cref{sec:source}) into a graph program
(\Cref{sec:target}), provided the source program is well formed (\Cref{sec:compiler-impl}).
Every call site becomes a node, every state a chain of state channels, and every loop a loop group.
We then prove in Rocq that compilation preserves weakest preconditions (\Cref{sec:verification}): every
specification proved for the source program with the logic of \Cref{sec:logic} holds for the compiled
graph, first under an abstract contract of graph execution, and from there under every executor of
\Cref{sec:target-exec}.

\subsection{Implementation}\label{sec:compiler-impl}

Our compiler translates a source program directly into a graph program, without an intermediate
representation. 
The compiler takes an expression and a \emph{context}, which records the channels available to the
expression (described below), and returns the nodes that compute the expression, the channel that carries
its value, and the context for the code that follows; we write
\[
C\vdash e\rightsquigarrow(G,c,C').
\]
The compiler is a function in a state monad that generates fresh channel names and may
fail; the judgment is its graph, with one rule per case of the source syntax. 

\emph{Contexts, copies, and triggers.} A context $C=\langle V,\Sigma,T\rangle$ records the channels that the rest of the program may still use.
$V$ maps every variable to the list of its \emph{unused copies}: a channel has at most one reader, so a
variable that is used $n$ times is bound to $n$ channels, made by one $\mathsf{Fanout}$ where the variable
is bound, and each use takes one of them.
$\Sigma$ maps every state to its current state channel, which carries its token.
$T$ is the list of unused copies of the scope's \emph{trigger}, a stream with one item per row of the
scope: one item at the top level, one per iteration in a loop body. Every node that runs once per row
without a data input consumes a copy of the trigger: the $\mathsf{Value}$ node of a constant or of a
loop's value, and the node of a call without argument.
How many copies are needed is computed from the syntax before compiling: $\mathsf{uses}(x,e)$ counts the
uses of $x$ in $e$, where a loop body that uses $x$ counts once, and $\mathsf{triggers}(e)$ counts the
consumers of the trigger (\Cref{app:copies}). We write $\mathsf{copies}(c,n)=(\bar c,G)$ for the $n$ copies $\bar c$ of the stream
on $c$ and the nodes $G$ that make them: none for $n\le 1$, one $\mathsf{Fanout}(c\to\bar c)$ otherwise.

\begin{figure}[p]
\centering
\footnotesize
\def\MathparLineskip{\lineskip=0.15cm}
\begin{subfigure}{\textwidth}
\centering
\begin{mathpar}
\inferrule*[right=Val]
  {T=t\cdot T'}
  {\langle V,\Sigma,T\rangle\vdash v\rightsquigarrow(\mathsf{Value}(v)(t\to o),\ o,\ \langle V,\Sigma,T'\rangle)}

\inferrule*[right=Var]
  {V(x)=c\cdot\bar c}
  {\langle V,\Sigma,T\rangle\vdash x\rightsquigarrow(\varepsilon,\ c,\ \langle V[x\mapsto\bar c],\Sigma,T\rangle)}

\inferrule*[right=Let]
  {C\vdash e_1\rightsquigarrow(G_1,c_1,C_1) \\
   \mathsf{copies}(c_1,\mathsf{uses}(x,e_2))=(\bar c,G_2) \\
   C_1[x\mapsto\bar c]\vdash e_2\rightsquigarrow(G_3,c,C_2)}
  {C\vdash\mathbf{let}\ x=e_1\ \mathbf{in}\ e_2\rightsquigarrow(G_1\,;G_2\,;G_3,\ c,\ C_2[x\mapsto C_1(x)])}

\inferrule*[right=Stateless Call]
  {C\vdash e\rightsquigarrow(G,c,C')}
  {C\vdash f(e)\rightsquigarrow(G\,;\mathsf{Stateless}(f)(c\to o),\ o,\ C')}

\inferrule*[right=Stateful Call]
  {C\vdash e\rightsquigarrow(G,c,\langle V,\Sigma,T\rangle) \\ \Sigma(s)=q}
  {C\vdash s.f(e)\rightsquigarrow(G\,;\mathsf{Stateful}(f)(q,c\to o,q'),\ o,\ \langle V,\Sigma[s\mapsto q'],T\rangle)}

\inferrule*[right=LetState]
  {C\vdash e_i\rightsquigarrow(G_i,c_i,C_1) \\ C_1[s\mapsto q]\vdash e_b\rightsquigarrow(G_b,c,C_2)}
  {C\vdash\mathbf{letstate}\ s=e_i\ \mathbf{in}\ e_b\rightsquigarrow
   (G_i\,;\mathsf{StateInit}(c_i\to q)\,;G_b,\ c,\ C_2[s\mapsto C_1(s)])}
\end{mathpar}
\caption{Constructs without loops.}
\label{fig:compile-base}
\end{subfigure}

\smallskip
\begin{subfigure}{\textwidth}
\centering
\begin{mathpar}
\inferrule*[right=For]
  {C\vdash e_i\rightsquigarrow(G_i,c_\ell,\langle V,\Sigma,t\cdot T\rangle) \\
   \mathsf{reuse}_{e_b}(V,k)=(V_b,V',G_r) \\\\
   \mathsf{copies}(c_x,\mathsf{uses}(x,e_b)+\mathsf{triggers}(e_b))=(\bar c_x\,\bar t_b,\ G_x) \\
   \langle V_b[x\mapsto\bar c_x],\Sigma,\bar t_b\rangle\vdash e_b\rightsquigarrow(G_b,\_,\langle\_,\Sigma',\_\rangle)}
  {C\vdash\mathbf{for}\ x\ \mathbf{in}\ e_i\ \mathbf{do}\ e_b\rightsquigarrow{} \\\\
   \bigl(G_i\,;\mathbf{loop}(\mathsf{For}(c_\ell\to c_x,k)\,;G_r\,;G_x\,;G_b)\,;\mathsf{Value}(())(t\to o),\
   o,\ \langle V',\Sigma',T\rangle\bigr)}
\end{mathpar}

\begin{tikzpicture}[scale=0.77,transform shape,x=0.88cm,y=1cm]
  \node[fnode] (gi) at (0,0) {$G_i$};
  \node[fnode] (for) at (2.3,0) {$\mathsf{For}$};
  \node[fnode,minimum height=0.65cm] (fx) at (5.4,0.7) {$\mathsf{Fanout}$};
  \node[fnode] (fk) at (4.8,-0.8) {$\mathsf{Fanout}$};
  \node[fnode] (re) at (6.9,-0.8) {$\mathsf{Reuse}$};
  \node[fnode,minimum width=1.3cm,minimum height=2.2cm] (gb) at (9.7,0) {$G_b$};
  \draw[fdata] (-1.4,0) node[flab,left]{outer trigger} -- (gi);
  \draw[fdata] (gi) -- node[flab,above]{$c_\ell$} (for);
  \draw[fdata] (for) |- node[flab,pos=0.75,above]{$c_x$ (elements)} (fx);
  \draw[fdata] (for) |- node[flab,pos=0.68,below]{$k$ (extents)} (fk);
  \draw[fdata] (fx.east |- 0,0.86) -- node[flab,above]{$x$} (gb.west |- 0,0.86);
  \draw[fdata] (fx.east |- 0,0.54) -- node[flab,below]{body trigger} (gb.west |- 0,0.54);
  \draw[fdata] (fk) -- node[flab,above]{$k_y$} (re);
  \draw[fdata] (-1.4,-1.6) node[flab,left]{$c_y$ (outer $y$)} -| (re.south);
  \draw[fdata] (re) -- node[flab,above]{$c'_y$} (gb.west |- re);
  \draw[fdang] (gb.east) -- ++(0.4,0) node[flab,right]{value discarded};
  \draw[fstate] (-1.4,1.7) node[flab,left]{$\Sigma(s)$} -- (9.7,1.7) -- (gb.north);
  \draw[fstate] (gb.south) -- (9.7,-1.9) -- (11.7,-1.9) node[flab,right]{$\Sigma'(s)$};
  \node[fnode] (val) at (2.3,-2.6) {$\mathsf{Value}(())$};
  \draw[fdata] (-1.4,-2.6) node[flab,left]{outer trigger} -- (val);
  \draw[fdata] (val) -- ++(1.6,0) node[flab,right]{$o$};
  \begin{scope}[on background layer]
    \node[fregion,fill=fLoop,fit=(for)(fx)(fk)(re)(gb)] (lg) {};
  \end{scope}
  \node[flab,anchor=south east] at (lg.north east) {$\mathbf{loop}$};
\end{tikzpicture}

\caption{Rule \textsc{For} and the graph it produces.}
\label{fig:compile-for}
\end{subfigure}

\smallskip
\begin{subfigure}{\textwidth}
\centering
\begin{mathpar}
\inferrule*[right=TrFix]
  {C\vdash e_i\rightsquigarrow(G_i,c_s,\langle V,\Sigma,T\rangle) \\
   \mathsf{copies}(a,\mathsf{uses}(x,e_b)+\mathsf{triggers}(e_b))=(\bar c_x\,\bar t_b,\ G_a) \\\\
   \langle[x\mapsto\bar c_x],[\bar s\mapsto\bar q_b],\bar t_b\rangle\vdash e_b\rightsquigarrow(G_b,c_f,\langle\_,\Sigma_b,\_\rangle)}
  {C\vdash\mathbf{trfix}\ x=e_i\ \mathbf{in}\ e_b\rightsquigarrow{} \\\\
   \bigl(G_i\,;\mathbf{loop}(\mathsf{Fix}(|\bar s|)(c_s,c_f,\Sigma(\bar s),\Sigma_b(\bar s)\to a,o,\bar q_b,\bar q_o)\,;G_a\,;G_b),\
   o,\ \langle V,\Sigma[\bar s\mapsto\bar q_o],T\rangle\bigr)}
\end{mathpar}

\begin{tikzpicture}[scale=0.77,transform shape,x=0.88cm,y=1cm]
  \node[fnode] (gi) at (0,0) {$G_i$};
  \node[fnode,minimum height=2.4cm,minimum width=1.0cm] (fix) at (2.6,0) {$\mathsf{Fix}$};
  \node[fnode,minimum height=0.65cm] (fo) at (5.0,0.5) {$\mathsf{Fanout}$};
  \node[fnode,minimum width=1.3cm,minimum height=2.4cm] (gb) at (8.4,0) {$G_b$};
  \draw[fdata] (-1.4,0) node[flab,left]{outer trigger} -- (gi);
  \draw[fdata] (gi) -- node[flab,above]{$c_s$} (fix.west |- 0,0);
  \draw[fdata] (fix.east |- 0,0.5) -- node[flab,above]{$a$} (fo);
  \draw[fdata] (fo.east |- 0,0.66) -- node[flab,above]{$x$} (gb.west |- 0,0.66);
  \draw[fdata] (fo.east |- 0,0.34) -- node[flab,below]{body trigger} (gb.west |- 0,0.34);
  \draw[fdata] (gb.south) -- (8.4,-1.7) -- node[flab,below]{$c_f$ (results of the rounds)} (2.9,-1.7)
    -- (fix.south -| 2.9,0);
  \draw[fdata] (fix.south -| 2.3,0) -- (2.3,-2.6) -- (10.6,-2.6) node[flab,right]{$o$};
  \draw[fstate] (-1.4,1.6) node[flab,left]{$\Sigma(s)$} -- (2.3,1.6) -- (fix.north -| 2.3,0);
  \draw[fstate] (fix.east |- 0,1.0) -- node[flab,above,pos=0.55]{$q_b$} (gb.west |- 0,1.0);
  \draw[fstate] (gb.north) -- (8.4,2.0) -- node[flab,above]{$\Sigma_b(s)$} (2.9,2.0) -- (fix.north -| 2.9,0);
  \draw[fstate] (fix.south -| 2.6,0) -- (2.6,-2.2) -- (10.6,-2.2) node[flab,right]{$q_o$};
  \begin{scope}[on background layer]
    \node[fregion,fill=fLoop,fit=(fix)(fo)(gb)] (lg) {};
  \end{scope}
  \node[flab,anchor=south east] at (lg.north east) {$\mathbf{loop}$};
\end{tikzpicture}

\caption{Rule \textsc{TrFix} and the graph it produces, for a body that calls one state $s$.}
\label{fig:compile-fix}
\end{subfigure}
\caption{Compilation rules; \Cref{app:compile-rules} adds those of $f()$ and $s.f()$. $o$, $q$, $q'$, $c_x$, $k$, $a$, $\bar q_b$, $\bar q_o$ are fresh channels;
$t\cdot T'$ is the list with first element $t$ and rest $T'$; $;$ concatenates node lists, in emission
order; a binding is restored when its scope ends. In \textsc{For}, $\bar c_x$ are the first $\mathsf{uses}(x,e_b)$ copies of the
elements and $\bar t_b$ the rest; $\mathsf{reuse}_{e_b}(V,k)=(V_b,V',G_r)$ repeats once per element, with a
$\mathsf{Reuse}(c_y,k_y\to c'_y)$ driven by a copy $k_y$ of the extents, every variable $y$ that $e_b$ uses.
In \textsc{TrFix}, $\bar s$ lists the states that $e_b$ calls, each once, and $\Sigma(\bar s)$ and
$\Sigma_b(\bar s)$ are their state channels before and after the body. Loop groups are shaded; in the
graphs, back edges go into $\mathsf{Fix}$.}
\label{fig:compile-rules}
\end{figure}
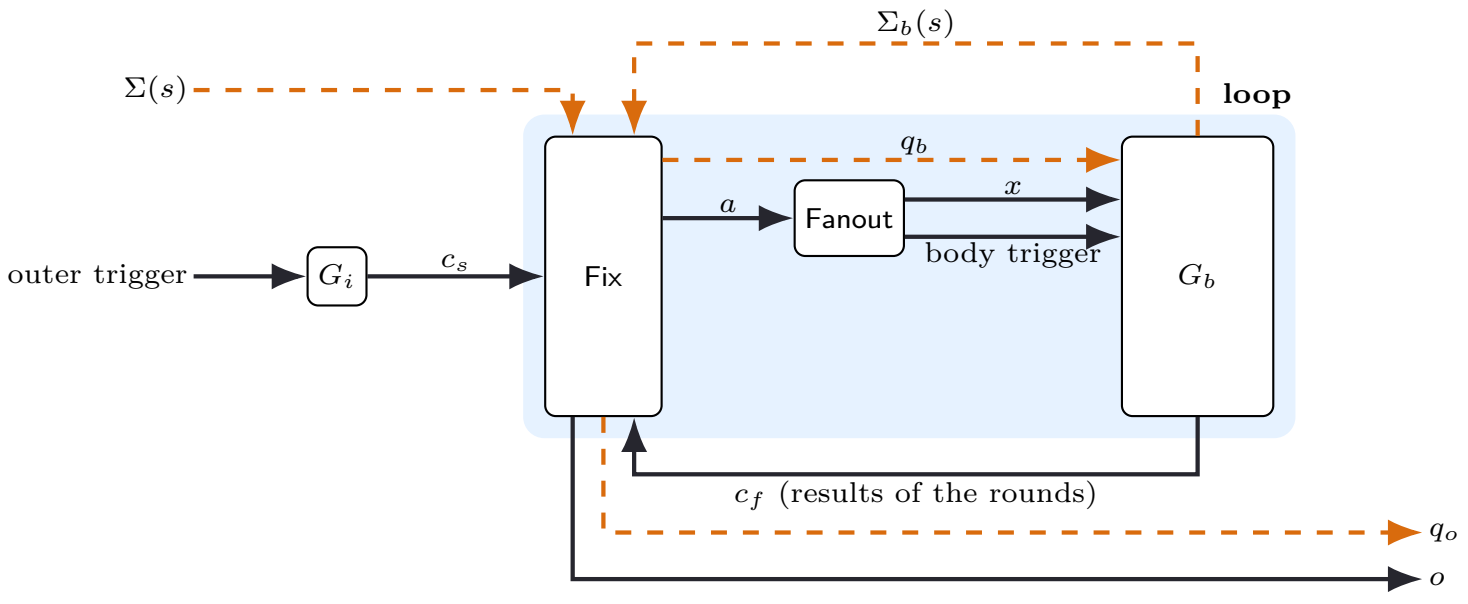

\emph{Compilation rules.} The constructs without loops (\Cref{fig:compile-base}) add at most one node each, plus the
$\mathsf{Fanout}$ nodes that make copies; the context passes the copies and the tokens along.
A value becomes a $\mathsf{Value}$ node that emits it once per item of a trigger copy, so once per row.
A variable compiles to no node at all: it takes one of its copies.
A \slet compiles its two parts in sequence and binds the variable to as many copies of the first part's
result as the second part uses.
A call compiles its argument and appends a call node that reads the argument's channel.
A stateful call also reads the token of its state, from $\Sigma(s)$, and writes it on a fresh channel
$q'$, which becomes the state's current state channel: successive calls on one state thus form the chain
of state channels of \Cref{sec:overview-compiler}.
A \sletstate appends a $\mathsf{StateInit}$ node that allocates the state with the value of its
initializer and emits the first token.

\emph{Loops.} A loop $\mathbf{for}\ x\ \mathbf{in}\ e_i\ \mathbf{do}\ e_b$ is compiled into a loop group
(\Cref{fig:compile-for}; rule \textsc{For}).
First, $e_i$ compiles to a channel $c_\ell$ that carries the lists to iterate over, and a node
$\mathsf{For}(c_\ell\to c_x,k)$ turns them into a stream of elements on $c_x$ and a stream of extents on $k$.
Every outer variable $y$ that the body uses gives up one copy $c_y$, and a node
$\mathsf{Reuse}(c_y,k_y\to c'_y)$, with a copy $k_y$ of the extents, repeats its value once per element;
the body uses copies of $c'_y$ instead of $c_y$.
The elements on $c_x$ are copied for the uses of $x$ in the body and for the body's trigger: every element
starts one row of the body.
The body is compiled in the context made of these copies, with the outer states, so that its stateful calls
continue the chains of state channels; the loop group collects the $\mathsf{For}$, the copies, the
$\mathsf{Reuse}$ nodes, and the body's nodes.
The value of the loop is $()$, emitted by a $\mathsf{Value}$ node on a copy of the outer trigger, once per
row of the outer scope.

A loop body must follow the \emph{loop-body rule}: it calls every state from at most one site, and
allocates no state.
A stateful node keeps the token of its state for its whole input stream, so a second site on the same
state could start only after the last iteration, and the calls would no longer follow the source order.
A state allocation needs one initial value, while a body has one row per iteration.

\emph{Tail recursion.} A tail recursion $\mathbf{trfix}\ x=e_i\ \mathbf{in}\ e_b$ is compiled into a loop group headed by a
$\mathsf{Fix}$ node (\Cref{fig:compile-fix}; rule \textsc{TrFix}).
First, $e_i$ compiles to the channel $c_s$ of the seeds, one per row of the scope.
$\mathsf{Fix}$ sends the argument of every round on a fresh channel $a$, whose items are copied for the
uses of $x$ in the body and for the body's trigger, so that the body runs one row per round.
The body is compiled in a context of its own: it sees only $x$, and every state $s$ that it calls
through a fresh channel $q_{b,s}$, on which $\mathsf{Fix}$ sends the token of $s$ at the start of every
round.
The body's result channel $c_f$ and its final state channels $\Sigma_b(\bar s)$ go back to
$\mathsf{Fix}$, which decides from the result whether to start another round (\Cref{sec:target-syntax});
these back edges are the cycles of compiled graphs.
After the last round of the last seed, $\mathsf{Fix}$ sends the tokens on $\bar q_o$, which become the
current state channels for the code that follows; the result of every seed goes on $o$.
Since the body cannot see the variables of the scope, $\mathsf{uses}(x,\cdot)$ and
$\mathsf{triggers}(\cdot)$ of a \strfix count only its initializer.

The two loops are compiled differently because they allow different parallelism.
The elements of a list are known as soon as the list arrives, so a $\mathsf{For}$ sends them all, and the
body of a \sfor works as a pipeline: while a node handles one element, the nodes before it already handle
the next ones.
To keep this pipeline, a stateful node holds the token of its state for the whole stream, and the
loop-body rule is the price.
The rounds of a \strfix cannot overlap, since the argument of a round is the result of the previous one.
With no pipeline to keep, the tokens travel with every round, through the back edges of $\mathsf{Fix}$: a
round runs from the states that the previous round left, as in the source, so the body may call a state
from several sites.
A \strfix may appear anywhere. Inside a \sfor body, its $\mathsf{Fix}$ receives one seed per iteration,
and it is, for the enclosing body, the single site of every state that it calls. Inside another round,
the outer round markers reach it on its seed channel, and it passes them on (\Cref{sec:target-syntax}).

\emph{Programs.} A closed program $e$ compiles to
\[
\begin{array}{@{}l@{}}
\mathbf{let}\ c_0=\mathbf{chan}\ \mathbf{in}\ \dots\ \mathbf{let}\ c_n=\mathbf{chan}\ \mathbf{in}\\
\quad\mathbf{execute}\bigl(\mathsf{Const}(())(\to t_0)\,;\mathsf{Fanout}(t_0\to\bar t)\,;G\,;\mathsf{Target}(c)\bigr),
\end{array}
\]
where $\langle\emptyset,\emptyset,\bar t\rangle\vdash e\rightsquigarrow(G,c,C')$, $\bar t$ are the
$\mathsf{triggers}(e)$ copies of the program's trigger, and $c_0,\dots,c_n$ are all channels the compilation
created.
The graph of \Cref{fig:overview-labyrinth-dfg} is the output of this compiler for Labyrinth, without the
trigger stream.

\emph{Well-formed programs.} The compiler accepts \emph{well-formed} programs (\Cref{app:wf}): every variable and every state is in scope, every
\sfor body follows the loop-body rule, and the body of every \strfix uses no variable but its argument
and allocates no state, since its states are those of the enclosing scope, passed round by
$\mathsf{Fix}$.

\begin{lemma}[Compilation succeeds, and compiled graphs are linear]\label{lem:compile-linear}
The compiler succeeds on every well-formed program. In every compiled graph, every channel has exactly one
writer and at most one reader. The writer of every channel comes before its reader in the emission order of
the nodes, except for the back edges of a $\mathsf{Fix}$, whose writers are nodes of the graph that come
after it.
\end{lemma}
\Cref{app:linear} defines the emission order and splits the lemma into the three results that prove
it.

\begin{corollary}[Least privilege]\label{cor:least-privilege}
In every compiled graph:
(i) a node holds only the channel ends it is created with, and no other node holds them;
(ii) a node can call at most one function, the $f$ of its kind;
(iii) the only nodes that can update a state are those that receive its token, one at a time, along the
chain of state channels of the state;
(iv) inside a tail recursion, this chain is a cycle through $\mathsf{Fix}$, which holds the token between
rounds.
\end{corollary}

The corollary follows from \Cref{lem:compile-linear} and from the node kinds of \Cref{fig:target-syntax}:
an operator reads and writes only the channels it is given, and its only other memory is its node state
and its local references (\Cref{sec:target-sem}).

\subsection{Preservation of Source Weakest Preconditions}\label{sec:verification}

We now show that compilation preserves weakest preconditions: every specification proved for a source
program with the logic of \Cref{sec:logic} holds for the compiled graph.
We first give graph programs a program logic in the same style, then state the result for an abstract
specification of graph execution, sketch its proof, and finally carry it to the executors of
\Cref{sec:target-exec}.

\emph{A program logic for graphs.} As for the source, the logic of graph programs is obtained from their \itree semantics by giving each part
of $E_{\mathit{dfg}}$ a logical handler:
\[
H_{\mathit{dfg}}\ \triangleq\ H_{\mathit{constr}}\oplus H_{\mathit{exec}}\oplus H_{\mathit{ub}}\oplus H_{\mathit{step}},
\]
and we write $\WPg{g}{\Phi}\triangleq\wpi{H_{\mathit{dfg}}}{\llbracket g\rrbracket}{\Phi}$.
$H_{\mathit{constr}}$ answers $\mathsf{chan}$ with a fresh channel and the ownership of its two ends, and
records the nodes created by $\mathsf{node}(n)$ (\Cref{app:graph-handlers}); $H_{\mathit{ub}}$ requires $\bot$, as in the source;
$H_{\mathit{step}}$ lets a step cost at most a later.
The execution event is the interesting one. Rather than fixing one executor, $H_{\mathit{exec}}$ specifies
execution abstractly:
\[
H_{\mathit{exec}}(\mathsf{execute}(\bar n),\Psi)\ \triangleq\ \WPt{\bar n}{\Psi},
\]
where $\WPt{\bar n}{\Psi}$, described next, is a contract between the graph $\bar n$ and whoever executes
it: the graph must establish it, and an executor may rely on it.
In short, the baton passes the calls and iterations in source order, all other firings are free, and the
graph never gets stuck.
The calls of the nodes are handled by the same function specifications as in the source, and the
iterations by the same handler $H_{\mathit{iter}}$.
A graph returns graph values, so a postcondition $\Phi$ of the source becomes
$\Phi^{\uparrow}(v)\triangleq\exists v'.\ v=\mathsf{val}(v')\ast\Phi(v')$, where $\mathsf{val}(v')$ is
the graph value that carries the source value $v'$.

\emph{The graph specification $\mathrm{WP}_{\mathsf{t}}$.} The baton of \Cref{sec:overview-preservation} is the heart of $\mathrm{WP}_{\mathsf{t}}$. Some firings must
happen in the order of the source program, the others are free.
The ordered ones are the \emph{control firings}: the calls; the firings of a $\mathsf{For}$ that take a
list or send a further element; and the firings of a $\mathsf{Fix}$ that take a seed or decide, after a
round, whether to continue.
The source program fixes one order of the control firings, its \emph{trace}, once the lists that the
$\mathsf{For}$ nodes take and the arguments of the rounds of every $\mathsf{Fix}$ are known: statements in
order, one row of a loop body per element, and one row of a tail-recursive body per round.
The rounds are revealed one at a time, since whether a round is followed by another one is known only
from its result; the trace is infinite if a tail recursion does not terminate.
All other firings (constants, copies, repetitions, and every firing on an end marker) may happen at any
time.

\begin{example}[A trace]\label[example]{ex:trace}
Except for its triggers and the value of the loop, the graph of \Cref{fig:target-example} is the
compiled form of
$\mathbf{letstate}\ s=0\ \mathbf{in}\ \mathbf{let}\ u=(\mathbf{for}\ x\ \mathbf{in}\ [1,2]\ \mathbf{do}\
s.\mathsf{add}(x))\ \mathbf{in}\ s.\mathsf{get}()$.
Its trace has five control firings: the $\mathsf{For}$ takes $[1,2]$ and sends $1$, an iteration; the call
$\mathsf{add}(1)$; the $\mathsf{For}$ sends $2$, an iteration; the call $\mathsf{add}(2)$; and the call
$\mathsf{get}()$.
The source-order run of \Cref{tab:target-exec} fires them in this order; the other runs fire the second
iteration before $\mathsf{add}(1)$.
\end{example}

Writing $\pi$ for a position in the trace, $\WPt{\bar n}{\Phi}$ (formally in \Cref{app:trace,app:wpt}) requires that there exist a predicate
$\mathsf{Ctl}$ on positions, the baton, and for every node $n$ an invariant $\mathit{Inv}_n$ and a
postcondition $\Psi_n$, such that:
\begin{enumerate}[label=(\roman*)]
\item \emph{Start.} $\mathsf{Ctl}(\pi_0)$ holds at the first position of the trace.
\item \emph{Calls.} The call handler of a node requires $\mathsf{Ctl}(\pi)$ at the node's own position
$\pi$. When the call returns an answer that the function's specification allows, the baton moves to the
next position.
\item \emph{Iterations.} At an iteration, $\mathsf{Ctl}(\pi)$ gives $H_{\mathit{iter}}(\mathsf{iter},\Psi)$,
with $\Psi$ the baton at the next position: the same obligation as in the source. The iterations are the
control firings that start a row of a body: a $\mathsf{For}$ that sends an element, and a $\mathsf{Fix}$
that takes a seed or decides to continue. A $\mathsf{For}$ that takes an empty list and a $\mathsf{Fix}$
that decides to stop move the baton without an iteration.
\item \emph{Nodes.} Every node meets its own specification: started in its initial state, every firing
preserves $\mathit{Inv}_n$, and when it is done it owns $\Psi_n$ of its value.
\item \emph{Progress.} In every reachable state, either the graph is done and so is the trace, or the node
at the baton's position can make its control firing, or some free firing can happen.
\item \emph{End.} When the baton is past the last position, the postconditions $\Psi_n$ of the nodes give
$\Phi$ of the graph's result.
\end{enumerate}
Progress is deadlock freedom. Without it, the weakest precondition alone would allow the graph to get stuck.
Persistent certificates record what the nodes know about the arguments of calls, the lists that a
$\mathsf{For}$ takes, and the decisions of a $\mathsf{Fix}$; node specifications produce them, and the
transitions of the baton consume them.

The intuition behind $\mathsf{Ctl}(\pi)$ is that it is \emph{the weakest precondition of the rest of the
source program from position $\pi$}.
Formally, $\mathsf{Ctl}$ is the greatest fixpoint of its own steps: at a position $\pi$, every transition
that $\mathrm{WP}_{\mathsf{t}}$ allows leads to $\mathsf{Ctl}$ at the next position, and the progress
obligation holds.
A proof of the source's weakest precondition is therefore used up one control firing at a time, exactly
when the graph makes that firing, and the obligation for the next one is built only then, from the answer
of the current call.

\begin{lemma}[Compilation preserves weakest preconditions]\label{lem:wp-pres}
Let $e$ be a well-formed program that compiles to the graph program $g$. For every postcondition $\Phi$
and every handler of iterations $H_{\mathit{iter}}$,
\[
\WPsrc{e}{\Phi}\ \vdash\ \WPg{g}{\Phi^{\uparrow}}.
\]
\end{lemma}

The lemma lets one reason about the sequential program and obtain a guarantee about its concurrent
execution: every specification that the source satisfies, the graph satisfies too.
The function specifications and $H_{\mathit{iter}}$ are shared by both sides, and $H_{\mathit{iter}}$ is
left open: every iteration of the source is matched by one iteration of the graph, so whatever an
iteration costs in the proof of the source, for instance a later, is paid in the same way by the graph.
The extra firings of the graph, such as those of $\mathsf{Fanout}$ and $\mathsf{Reuse}$ nodes, cost
nothing, since a later on the target only guards a continuation that the proof already owns
($P\vdash\triangleright P$).
The lemma goes in one direction only: it does not state that the graph has every behaviour of the
source.

We prove the lemma in Rocq for every well-formed program, tail recursions included, and without admitted
lemmas.
Adequacy holds for each executor: the theorems at the end of this section carry the lemma to the
weakest precondition of every executor's run, which is adequate.

\begin{proof}[Proof sketch of \Cref{lem:wp-pres}]
The proof splits at $\mathrm{WP}_{\mathsf{t}}$ into two independent parts.

\emph{(I) Compiled graphs meet $\mathrm{WP}_{\mathsf{t}}$.}
We take for $\mathsf{Ctl}$ the greatest fixpoint of clauses (ii), (iii), (v) and (vi), so these hold by
construction, and only (i) and (iv) need a proof.
Clause (iv) follows from the linearity of compiled graphs (\Cref{lem:compile-linear}).
Clause (i) follows from a \emph{control lemma} (\Cref{lem:control}), by induction on the compilation
rules of \Cref{fig:compile-rules}. 
Its proof walks the program from front to back, \emph{consuming} $\WPsrc{e}{\Phi}$
one event at a time, while $\mathsf{Ctl}$ at each position is obtained from $\mathsf{Ctl}$
at the next one, back from the end, through continuations;
$\mathsf{Ctl}$ at the first position of the trace is what we get once
the whole chain is assembled.

\emph{(II) The source-order scheduler meets $\mathrm{WP}_{\mathsf{t}}$.} It runs every graph $\bar n$ that
satisfies $\WPt{\bar n}{\Phi}$ to a result that satisfies $\Phi$, under $H_{\mathit{src}}$.
Its invariant keeps the baton at the next control firing, and the progress clause ensures that some node
can always fire, so no $\mathsf{recv}$ finds an empty channel.
Part (II) does not depend on the compiler.
\end{proof}

\emph{Preservation for every executor.} We now carry the lemma to the executors of \Cref{sec:target-exec}, as weakest preconditions of their runs
under $H_{\mathit{src}}$, since a run raises the events of the source: calls, iterations, undefined
behaviour, and steps.
We write $\WPx{x}{g}{\Phi}$ for the weakest precondition of the run of $g$ by the executor $x$; $x$ is
$\mathit{order}$ for the source-order scheduler, $\mathit{demonic}$ for the demonic scheduler, and
$\mathit{local}$ and $\mathit{par}$ for the local and the parallel executor (\Cref{app:wpx}).
The three theorems below assume that the handler of $\mathsf{step}$ events guards the continuation of a
step by a later; \Cref{app:theorems} states them with all premises, as checked in Rocq.

\begin{theorem}[Preservation in source order]\label{thm:wp-pres-order}
Let $e$ be a well-formed program that compiles to $g$. For every postcondition $\Phi$ and every handler of
iterations $H_{\mathit{iter}}$,
$\WPsrc{e}{\Phi}\vdash\WPx{order}{g}{\Phi^{\uparrow}}$.
\end{theorem}

The theorem composes (I) and (II) at $\mathrm{WP}_{\mathsf{t}}$.
The other executors need unguarded iterations.
A source proof by L\"ob induction strips the later of an iteration with the graph's step at that
iteration.
Only the source-order scheduler fires an iteration when the baton is there; the other executors may fire
it earlier, and the baton catches up by a ghost update, which takes no step.

\begin{theorem}[Preservation under every schedule]\label{thm:wp-pres-sched}
Let $e$ be a well-formed program that compiles to $g$. For every postcondition $\Phi$,
$\WPsrctot{e}{\Phi}\vdash\WPx{demonic}{g}{\Phi^{\uparrow}}$.
\end{theorem}

The choice of the scheduler is universal in the weakest precondition (a demonic choice): every schedule
satisfies what the source satisfies.
The proof reproves (II) for the demonic scheduler, in which nodes may make control firings ahead of the
baton.
To allow this, the baton of $\mathrm{WP}_{\mathsf{t}}$ may lag behind the nodes, and it catches up with each
of them by a transition of $\mathrm{WP}_{\mathsf{t}}$; the scheduler keeps the run's position in the source
order only for this proof.

\begin{theorem}[Preservation without a scheduler]\label{thm:wp-pres-distr}
Let $e$ be a well-formed program that compiles to $g$. For every postcondition $\Phi$,
$\WPsrctot{e}{\Phi}\vdash\WPx{local}{g}{\Phi^{\uparrow}}$ and
$\WPsrctot{e}{\Phi}\vdash\WPx{par}{g}{\Phi^{\uparrow}}$.
\end{theorem}

The local executor is the demonic scheduler without the position that the scheduler keeps for the proof:
their runs are equal up to silent steps, so \Cref{thm:wp-pres-sched} carries over.
A round of the parallel executor fires nodes with disjoint inputs, and has the effect of firing them one
after another; every parallel run of a compiled graph is therefore one of the runs of the local executor,
whose weakest precondition covers all of them.

\emph{Adequacy.} The logic of the source is adequate for unguarded iterations (\Cref{app:src-wp}): given an implementation
of the functions that is proved to meet their specifications, $\WPsrctot{e}{\varphi}$ with a pure
postcondition $\varphi$ is a fact about the program's runs.
With the theorems above, this applies to the compiled graph, from a proof of $\WPsrctot{e}{\varphi}$.
Run by any of the executors from the empty heap, with the calls answered by the implementation and a
budget of steps, the graph never reaches undefined behaviour and, if it returns before the budget runs
out, returns a value that satisfies $\varphi$.
This is partial correctness, by design: a source program may diverge, and so may its graph.
For Labyrinth, we instantiate the theorems in Rocq with no hypotheses left, from the proof of
$\WPsrctot{\mathit{labyrinth}}{\Phi}$ (\Cref{ex:labyrinth-verification}) and a verified implementation of
its functions; every run of its compiled graph, in parallel or in source order, never reaches undefined
behaviour and, if it returns within its budget, returns pairwise disjoint routes.
Together, \Cref{lem:wp-pres,thm:wp-pres-order,thm:wp-pres-sched,thm:wp-pres-distr} give the end-to-end
result: a weakest precondition proved for the sequential source program holds for its compiled graph
under every executor of \Cref{sec:target-exec}.

\emph{Mechanization.} All definitions and results of
\Cref{sec:source,sec:target,sec:compiler} are mechanized in Rocq with Iris and the ITrees library, in about 70k lines of specifications and proofs, except
\Cref{cor:least-privilege}, which we argue informally; so is Labyrinth (\Cref{sec:overview}), with both
proofs of \Cref{ex:labyrinth-verification}, a verified implementation of its functions, and the
instantiation of our theorems for its compiled graph.
No result depends on an admitted lemma; the only axioms of the headline theorems are uniqueness of
identity proofs and the ITrees axiom that bisimilar trees are equal.
\Cref{app:theorems} states the theorems as checked in Rocq, and the web page of the artifact links each
result to its source.

\section{Related Work}\label{sec:relatedW}

\emph{Compilers from sequential programs to dataflow} already exist; our contribution is a mechanized
proof that one preserves the source specifications.
ConDRust~\cite{condrust}, from which Labyrinth comes, compiles sequential Rust programs to deterministic
graphs of threads and FIFO channels, and argues informally that compilation preserves semantics and
determinism.
Cascade~\cite{cascade} targets a streaming dataflow system, Ebcioglu and San~\cite{multifpga} multi-FPGA
accelerators, and TileLoom~\cite{tileloom} spatial dataflow accelerators; their correctness rests on
construction and testing.

Two lines of work establish the correctness of compilation to dataflow for hardware, by \emph{validation}
and by \emph{verification}; both relate final memories or traces, while we relate specifications.
FlowCert~\cite{flowcert} validates each run in two steps: a simulation on a \emph{canonical schedule} that
follows the source, and permission tokens, found by an SMT solver, that make every schedule reach the same
final memory.
Our proof has the same two steps (\Cref{thm:wp-pres-order,thm:wp-pres-sched}), with the tokens of the
states in the role of the permissions.
Lin et al.~\cite{letitflow} verify the core passes of such a compiler in Lean; like us, they treat calls as
uninterpreted events.
Their operators share a memory, and capability types decide which accesses may run in parallel, so they
pipeline every loop, while we pipeline \sfor loops only, since the rounds of a \strfix depend on each other.
We relate separation-logic specifications instead: every weakest precondition of the source holds for
the graph, also for programs that call external functions known only through their specifications, and
in source order also for partial-correctness proofs by L\"ob induction (\Cref{thm:wp-pres-order});
our contract of graph execution includes deadlock freedom and holds for every executor that implements
it, among them execution without a scheduler (\Cref{thm:wp-pres-distr}).
Bell~\cite{bell-cpp13} proves sound in Coq loop-parallelizing transformations such as decoupled software
pipelining, which splits a loop into stages connected by queues, for a process calculus and with respect
to behaviours.
V\'elus~\cite{velus} verifies compilation in the opposite direction, from the synchronous dataflow language
Lustre to sequential code.
Beyond dataflow, CompCertTSO~\cite{compcerttso} verifies compilation to a concurrent, relaxed-memory target;
its correctness, like our \Cref{lem:wp-pres}, goes in one direction, but is stated as inclusion of traces.

Our executors follow the two classic views of \emph{dataflow execution} (\Cref{sec:target-exec}): the local and
the parallel executor fire nodes as soon as their inputs hold data, without a scheduler, and the
schedulers choose which node fires next.
In a Kahn process network~\cite{kahn74}, processes block on their inputs, and the histories of the channels
are a least fixpoint, independent of the order of execution; in a dataflow process
network~\cite{lee-parks-dpn}, actors fire when their firing rules hold, and Lee and
Matsikoudis~\cite{lee-matsikoudis-firing} relate the two views.
Our proofs need no such connection, since all our executors are defined by firings.
Flo~\cite{flo} identifies determinism under any schedule and \emph{streaming progress} as the properties a
streaming program should have, and proves on paper that composition preserves them; the progress clause of
$\mathrm{WP}_{\mathsf{t}}$ is a property of this kind, deadlock freedom, proved mechanically for the graph
of every well-formed program.

\emph{Multitier programming} and \emph{automatic partitioning} also split one program across locations, but they decide
\emph{where} code runs, not which parts of it can run in parallel.
Neubauer and Thiemann~\cite{seqToMulti} come closest to our setting: they transform a sequential program with
location annotations into one process per tier, communicating over session-typed channels, and prove on paper
that every step yields a weakly bisimilar program.
\emph{Choreographic programming}~\cite{chor} also projects a global program, a choreography of the
interactions between participants, onto one process per participant,
and the projection has been mechanized for core choreographies~\cite{cruzfilipe-ictac21}, for the
higher-order Pirouette~\cite{pirouette}, and end to end down to machine code in Kalas~\cite{kalas}.
In these works the programmer places every computation on a participant, and the result relates the
processes to the choreography operationally; our compiler chooses the partition itself and preserves
specifications.
Multitier languages, from Links~\cite{links}, Ur/Web~\cite{urweb}, and Hop.js~\cite{hopjs} to
ScalaLoci~\cite{scalaloci}, let one program describe client, server, and database, and generate the
communication between them~\cite{multitier-survey}; ScalaLoci has a mechanized core calculus, which proves
type soundness and static placement.
Coign~\cite{coign}, J-Orchestra~\cite{jorchestra}, and Pyxis~\cite{pyxis} partition existing applications
automatically, choosing the placement from profiles of their communication.
Their units of distribution are tiers, components, or statements connected by remote calls; ours are the
nodes of a dataflow graph, and what we preserve is specifications rather than an equivalence.

\emph{Secure program partitioning} compiles one sequential program into a distributed one that preserves its
security properties against an adversary.
Jif/split~\cite{jifsplit} and Swift~\cite{swift} place the code and data of a program with
information-flow labels on hosts that the labels trust, and Viaduct~\cite{viaduct} chooses cryptographic
protocols where no host is trusted enough.
Acay et al.~\cite{cryptosyn} prove, on paper and for a core of Viaduct's language without loops, that each
step of this compilation, which goes through choreographies, is a simulation in the style of universal composability, which implies robust
hyperproperty preservation~\cite{beyondFA}.
Where this line of work preserves security, we preserve functional specifications, with a mechanized proof,
for a language with loops and state.

\emph{Compartmentalization} splits a program into compartments that hold only the privileges they need.
Capability machines such as CHERI~\cite{7163016} support it in hardware, with compartments that interact
through synchronous cross-compartment calls.
ACES~\cite{aces} and EC~\cite{ecSP23} partition embedded programs into compartments automatically, by
program analysis, without a proof that the partitioned program behaves as the original.
SECOMP~\cite{secomp} is a formally secure compiler for C programs whose compartments the developer
declares; compartments interact by calls and returns and exchange only scalar values.
Our compiler is a verified automatic compartmentalization: it partitions a sequential program into nodes that
share no channel end, call at most one function each, and update a state only while they hold its token (\Cref{cor:least-privilege}), and the partitioned
program provably satisfies every specification of the original, also when its nodes run in parallel
without a scheduler (\Cref{thm:wp-pres-distr}).
Unlike SECOMP, however, our guarantee is functional: it assumes that every node meets its own specification,
and does not protect against compromised compartments.

Several approaches to \emph{distributed cloud programming} also treat sequential code as the source of truth for a distributed
system.
Durable Functions~\cite{durablefunctions} gives a stateful serverless programming model a formal semantics,
and proves that its runtime executes workflows observably exactly once.
ECROs~\cite{ecros} turn a sequential data type and a distributed specification into a replicated object; an
SMT solver finds which calls commute and need no coordination, an intuition that \Cref{thm:wp-pres-sched}
shares.
Service Weaver~\cite{serviceweaver} and Monolift~\cite{monolift} let developers write a logical monolith
whose components the runtime or the compiler deploys as services.
In the Hydro stack~\cite{hydrolysis,queryrewrites}, single-node dataflow programs are distributed by rewrites
that are correct by construction, applied by hand.
These approaches establish properties of a runtime or of individual rewrites; we prove that a compiler
preserves the specifications of the sequential program.

Our semantics and program logics build on \emph{interaction trees}.
We obtain the program logics of both languages from logical handlers, following Vistrup et
al.~\cite{plalacarte}.
Choice trees~\cite{chtrees} extend interaction trees with internal nondeterminism, which suits concurrent
systems; we model the choices of the scheduler with an event instead, so that the program logics of
\Cref{sec:logic} apply unchanged.
Conditional contextual refinement~\cite{ccr} combines refinement with separation-logic specifications,
for the modules of a sequential program.
Where certificate translation~\cite{certtrans} carries a proof through an optimizing compiler as a
certificate for the compiled program, Abate et al.~\cite{wppres-arxiv} preserve weakest preconditions, so
that the proof carries over unchanged.
A concurrent target needs more: a specification of graph execution, $\mathrm{WP}_{\mathsf{t}}$, whose
baton orders the calls and iterations while the other firings are free; a progress obligation, which rules
out deadlocks; and, for executors other than the source-order scheduler, control firings ahead of the
baton.

\section{Conclusions and Future Work}\label{sec:conclusions}

We presented a compiler from a sequential, imperative language with state, loops and tail recursion to a
dataflow language, verified in the Rocq proof assistant.
Both languages are given a formal semantics with interaction trees and weakest precondition calculi derived from it. 
Our compiler preserves weakest preconditions: every specification proved for a sequential program holds for
its compiled graph.
The result is stated against an abstract specification of graph execution, $\mathrm{WP}_{\mathsf{t}}$,
which our executors implement or inherit (\Cref{thm:wp-pres-order,thm:wp-pres-sched,thm:wp-pres-distr}),
and so holds for every other executor that implements it.
Developers can thus reason about the sequential program alone, as we did for Labyrinth
(\Cref{sec:overview}), and obtain partial correctness for every run of its compiled graph, under the
restriction on iterations discussed below.

\smallskip
Our source language is inspired by a core subset of Rust; extending it towards a larger subset of Rust
is future work.

\smallskip
Our compiler exposes task and pipeline parallelism: calls that neither depend on each other nor share a
state run at the same time, and the iterations of a \sfor overlap in a pipeline.
The next step is data parallelism, which would process several elements of a list at once by replicating a
node.
A stateless call in a loop body could be replicated, with its stream split among the replicas and their
results merged back in order; a stateful call could not, since the token of its state orders its calls.
We leave such replication, and its proof, as future work.

\smallskip
While \Cref{thm:wp-pres-order} allows for proofs by L\"ob induction, the theorems that carry weakest
preconditions to the other executors, \Cref{thm:wp-pres-sched} for the demonic scheduler and
\Cref{thm:wp-pres-distr} for execution without any scheduler, need unguarded iterations: they accept source proofs in which every \strfix comes with a decreasing measure.
The reason is that these executors may fire an iteration before the baton reaches it, so the graph's step
at that iteration comes too early to strip the later that a proof by L\"ob induction puts there.
In future work, we plan to use later credits~\cite{SpiesGTJKBD22}, which let a step yield a credit that
can strip a later at any time after the step: an iteration that fires early would keep its credit until the baton
catches up with it.

\smallskip
Beyond our executors, we plan to support the backends of~\Cref{fig:intro-goal}: a multi-core runtime, a cluster, and a
serverless platform.
Each backend that implements $\mathrm{WP}_{\mathsf{t}}$ inherits our result.
This is an obligation on the backend alone, independent of the compiler, as the proofs for our two
schedulers show.

\smallskip
Since our compiler is a verified automatic compartmentalization (\Cref{sec:relatedW}), each node of a
compiled graph can run in its own compartment of a capability machine such as CHERI.
Protecting the specification against compromised compartments requires a robust variant of weakest
precondition preservation~\cite{beyondFA}, which we leave as future work.

\section*{AI Usage Disclosure}

We developed the Rocq mechanization with the help of AI assistants: Claude Code (Anthropic) and ChatGPT
(OpenAI).
The authors designed the languages, their semantics and program logics, the compiler and the graph
specification $\mathrm{WP}_{\mathsf{t}}$, and set the statements of the main results; under the authors'
direction and review, the assistants wrote definitions, auxiliary lemmas and their statements, and proof
scripts.
We also used these assistants to help with writing this paper.
The authors take responsibility for the whole work.

\bibliography{references}

\clearpage
\appendix
\section{Appendix: Definitions Omitted from the Main Text}\label{app:defs}

This appendix collects the definitions that the main text gives only informally, in the order in which
the main text uses them.
The web page of the artifact (contribution 4 in \Cref{sec:intro}) links each of them to its Rocq
source.

\subsection{Source Language}\label{app:source}

\subsubsection{Denotation}\label{app:src-sem}
The results of a \strfix round are values: $\mathsf{Done}(v)\triangleq(0,v)$ and
$\mathsf{Continue}(v)\triangleq(1,v)$.
The denotation $\llbracket e\rrbracket_{\sigma,\rho}$ of \Cref{sec:source-sem} is
{\small
\[
\begin{array}{@{}r@{\ }c@{\ }l@{}}
\llbracket v\rrbracket_{\sigma,\rho} &=& \mathsf{Ret}\ v\qquad
\llbracket x\rrbracket_{\sigma,\rho}\ =\ \mathsf{Ret}\ \rho(x)\\
\llbracket\mathbf{let}\ x=e_1\ \mathbf{in}\ e_2\rrbracket_{\sigma,\rho} &=& v\leftarrow\llbracket e_1\rrbracket_{\sigma,\rho};\
  \llbracket e_2\rrbracket_{\sigma,\rho[x\mapsto v]}\\
\llbracket f()\rrbracket_{\sigma,\rho} &=& \mathsf{slcall}(f,[\,])\qquad
\llbracket f(e)\rrbracket_{\sigma,\rho}\ =\ v\leftarrow\llbracket e\rrbracket_{\sigma,\rho};\ \mathsf{slcall}(f,[v])\\
\llbracket s.f()\rrbracket_{\sigma,\rho} &=& \mathsf{sfcall}(f,\sigma(s),[\,])\qquad
\llbracket s.f(e)\rrbracket_{\sigma,\rho}\ =\ v\leftarrow\llbracket e\rrbracket_{\sigma,\rho};\
  \mathsf{sfcall}(f,\sigma(s),[v])\\
\llbracket\mathbf{letstate}\ s=e_i\ \mathbf{in}\ e_b\rrbracket_{\sigma,\rho} &=& v\leftarrow\llbracket e_i\rrbracket_{\sigma,\rho};\
  \ell\leftarrow\mathsf{alloc}(v);\ \llbracket e_b\rrbracket_{\sigma[s\mapsto\ell],\rho}\\
\llbracket\mathbf{for}\ x\ \mathbf{in}\ e_i\ \mathbf{do}\ e_b\rrbracket_{\sigma,\rho} &=&
  [v_1,\dots,v_n]\leftarrow\llbracket e_i\rrbracket_{\sigma,\rho};\\
  && \mathsf{iter};\ \llbracket e_b\rrbracket_{\sigma,\rho[x\mapsto v_1]};\ \dots\
  \mathsf{iter};\ \llbracket e_b\rrbracket_{\sigma,\rho[x\mapsto v_n]};\ \mathsf{Ret}\ ()\\
\llbracket\mathbf{trfix}\ x=e_i\ \mathbf{in}\ e_b\rrbracket_{\sigma,\rho} &=& v_0\leftarrow\llbracket e_i\rrbracket_{\sigma,\rho};\
\\ && \mathsf{loop}\ \bigl(\lambda v.\ \mathsf{iter};\ r\leftarrow\llbracket e_b\rrbracket_{\sigma,[x\mapsto v]};\
  \mathsf{Ret}\ r\bigr)\ v_0,
\end{array}
\]
}
where $\mathsf{loop}$ continues with $v'$ on $r=\mathsf{Continue}(v')$ and returns $v'$ on
$r=\mathsf{Done}(v')$.
The tree raises $\mathsf{ub}$ when $\rho(x)$ or $\sigma(s)$ is undefined, when the iterable of a \sfor is
not a list, and when a round returns a value that is neither $\mathsf{Done}(v')$ nor
$\mathsf{Continue}(v')$; a stateful call looks up $\sigma(s)$ before it evaluates its argument.

\subsubsection{Function specifications and handlers}\label{app:src-handlers}
A program is verified against a family of function specifications: for every function name $f$, a
predicate $\mathsf{spec}_f(\bar v,r)$ for a stateless function and $\mathsf{spec}_f(w,\bar v,w',r)$ for a
stateful one.
Specifications are persistent, since they are facts about values: a graph copies the result of a call to
every reader, and each reader needs the facts about it.
The handler of the program heap is the standard one of separation logic:
\[
\begin{array}{@{}r@{\ }c@{\ }l@{}}
H_{\mathit{heap}}(\mathsf{alloc}(v),\Psi) &\triangleq& \forall\ell.\ \ell\mapsto v\mathrel{-\!\!*}\Psi(\ell)\\
H_{\mathit{heap}}(\mathsf{load}(\ell),\Psi) &\triangleq& \exists w.\ \ell\mapsto w\ast(\ell\mapsto w\mathrel{-\!\!*}\Psi(w))\\
H_{\mathit{heap}}(\mathsf{store}(\ell,v),\Psi) &\triangleq& \exists w.\ \ell\mapsto w\ast(\ell\mapsto v\mathrel{-\!\!*}\Psi(())).
\end{array}
\]
Together with $H_{\mathit{fun}}$ and $H_{\mathit{ub}}$ of \Cref{sec:logic}, and the handler of
$\mathsf{step}$ events of \Cref{app:graph-handlers}, which the source never raises, it forms the handler
$H_{\mathit{node}}$ of the events that the source shares with graphs.

\subsubsection{Weakest preconditions and adequacy}\label{app:src-wp}
The points-to assertions $\ell\mapsto w$ live in an instance $G$ of the ghost state of the program heap,
and the source weakest precondition holds for every instance:
\[
\WPsrc{e}{\Phi}\ \triangleq\ \forall G.\ \wpi{H_{\mathit{node}}^{G}\oplus H_{\mathit{iter}}}{\llbracket e\rrbracket}{\Phi},
\]
where $H_{\mathit{node}}^{G}$ interprets the program heap in $G$.
A proof never depends on the instance, so the quantifier costs nothing; it lets the proof of
\Cref{lem:wp-pres} run the source on a heap of its own.

An \emph{implementation} $\mathit{ctx}$ answers every call by an \itree over the events of the program
heap and of undefined behaviour; it \emph{meets the specifications} when, in the program logic,
\[
\begin{array}{@{}l@{}}
\Box\bigl(\forall f,\bar v.\ \wpi{H}{\mathit{ctx}(\mathsf{slcall}(f,\bar v))}{r.\ \mathsf{spec}_f(\bar v,r)}\bigr)
\quad\text{and}\\
\Box\bigl(\forall f,\ell,w,\bar v.\ \ell\mapsto w\mathrel{-\!\!*}\wpi{H}{\mathit{ctx}(\mathsf{sfcall}(f,\ell,\bar v))}
  {r.\ \exists w'.\ \ell\mapsto w'\ast\mathsf{spec}_f(w,\bar v,w',r)}\bigr),
\end{array}
\]
with $H$ the handler of its events.
The logic is adequate for unguarded iterations: let $t$ be an \itree over the events of the source and
$\mathsf{step}$, $\varphi$ a pure postcondition, and $\mathit{ctx}$ an implementation that meets the
specifications, such that $\wpi{H_{\mathit{node}}^{G}\oplus H^{\mathit{free}}_{\mathit{iter}}}{t}{v.\,\varphi(v)}$
for every instance $G$.
Then running $t$ with the calls answered by $\mathit{ctx}$, from the empty program heap, with a budget of
steps when $\mathsf{step}$ costs a later, is equivalent up to $\mathsf{Tau}$ steps to a run that returns a
value $v$ with $\varphi(v)$ or exhausts its budget; it never reaches undefined behaviour.
The theorems of \Cref{app:theorems} turn this result into facts about the runs of compiled graphs.

\subsection{Target Language}\label{app:target}

\subsubsection{Streams}\label{app:streams}
Channels carry ordinary values, and the stream protocol is a convention of the operators, encoded with a
tag in the first component of a pair:
\[
\mathsf{item}(v)\triangleq(0,v),\qquad
\mathsf{end}\triangleq(1,()),\qquad
\mathsf{round}\triangleq(2,()).
\]
A data channel carries items followed by one end marker; inside the body of a $\mathsf{Fix}$ loop, round
markers split its items into rounds (\Cref{sec:target-syntax}).
A state channel carries the token of a state, a location $\ell$ of the program heap, without a tag.
An operator that receives a value with another tag where it expects a stream value raises $\mathsf{ub}$.

\subsubsection{Operator language}\label{app:op-lang}
The full syntax of operators is
\[
\begin{array}{@{}r@{\ }c@{\ }l@{}}
o &::=& x \mid v \mid \mathbf{let}\ x=o\ \mathbf{in}\ o \mid o;o \mid \mathbf{if}\ o\ \mathbf{then}\ o\
  \mathbf{else}\ o \mid o\odot o \mid (o,o) \mid \mathbf{fst}\ o \mid \mathbf{snd}\ o\\
  &\mid& \mathbf{isnil}\ o \mid \mathbf{head}\ o \mid \mathbf{tail}\ o \mid \mathbf{length}\ o\\
  &\mid& \mathbf{recv}(c) \mid \mathbf{send}(c,o) \mid f(\bar o) \mid o.f(\bar o) \mid \mathbf{newstate}(o)
  \mid \mathbf{ref}\ o \mid\ !o \mid o:=o,
\end{array}
\]
where $\odot$ is an arithmetic operator and a condition holds when it is a non-zero integer.
The denotation of an operator is an \itree over $E_{\mathit{op}}$ (\Cref{sec:target-sem}), given by one
equation per construct, up to $\approx$; a condition that is not an integer, and $\mathbf{fst}$,
$\mathbf{snd}$, $\mathbf{head}$ or $\mathbf{tail}$ of a value of the wrong shape, raise $\mathsf{ub}$:
\[
\begin{array}{@{}r@{\ }c@{\ }l@{}}
\llbracket\mathbf{let}\ x=o_1\ \mathbf{in}\ o_2\rrbracket &=& v\leftarrow\llbracket o_1\rrbracket;\
  \llbracket o_2[v/x]\rrbracket\\
\llbracket\mathbf{if}\ o\ \mathbf{then}\ o_1\ \mathbf{else}\ o_2\rrbracket &=& n\leftarrow\llbracket o\rrbracket;\
  \text{if } n\neq 0 \text{ then } \llbracket o_1\rrbracket \text{ else } \llbracket o_2\rrbracket\\
\llbracket\mathbf{fst}\ o\rrbracket &=& (a,\_)\leftarrow\llbracket o\rrbracket;\ \mathsf{Ret}\ a\\
\llbracket\mathbf{recv}(c)\rrbracket &=& \mathsf{recv}(c)\\
\llbracket\mathbf{send}(c,o)\rrbracket &=& v\leftarrow\llbracket o\rrbracket;\ \mathsf{send}(c,v);\ \mathsf{Ret}\ ()\\
\llbracket f(\bar o)\rrbracket &=& \bar v\leftarrow\llbracket\bar o\rrbracket;\ \mathsf{slcall}(f,\bar v)\\
\llbracket o.f(\bar o)\rrbracket &=& \ell\leftarrow\llbracket o\rrbracket;\ \bar v\leftarrow\llbracket\bar o\rrbracket;\
  \mathsf{sfcall}(f,\ell,\bar v)\\
\llbracket\mathbf{newstate}(o)\rrbracket &=& v\leftarrow\llbracket o\rrbracket;\ \mathsf{alloc}(v).
\end{array}
\]
The local references $\mathbf{ref}$, $!$ and $:=$ act on the operator heap, which a state handler
interprets inside the denotation of each activation, starting from the empty heap; they raise no event of
$E_{\mathit{op}}$, and nothing an activation stores in them survives it: the node state is the only memory
of a node between activations.
The communication events are handled by the executor on the communication heap, a map from channels to
queues: $\mathsf{send}(c,v)$ appends $v$ to the queue of $c$, and $\mathsf{recv}(c)$ removes its first
value, or raises $\mathsf{ub}$ if the queue is empty or $c$ was never allocated.

\subsubsection{Operators of the node kinds}\label{app:operators}
An activation runs the operator of its node once on the node state $s$ and returns
$\mathsf{Done}(v)$, $\mathsf{Continue}(s')$, or $\mathsf{Iterate}(s')$; the last one is returned by
$\mathsf{For}$ and $\mathsf{Fix}$ when the activation starts an iteration, and is otherwise treated as
$\mathsf{Continue}(s')$.
We write
\[
\mathbf{on}\ c\ \{\ x\Rightarrow o_1\mid\mathsf{end}\Rightarrow o_2\mid\mathsf{round}\Rightarrow o_3\ \}
\]
for receiving one value from $c$ and running $o_1$ with $x$ bound to the content of an item, $o_2$ on the
end marker and $o_3$ on the round marker; any other value raises $\mathsf{ub}$, and so does a missing
case.
$\bot$ is the node state of a node that holds nothing yet.
Every node reads, in a given node state, a fixed set of inputs: those that its operator receives from in
that state.
The operators below show the channels in the order in which the node kind takes them.
{\small
\[
\begin{array}{@{}l@{}}
\mathsf{Const}(v)(\to o),\ s_0=0:\\
\quad\mathbf{if}\ s=0\ \mathbf{then}\ \mathbf{send}(o,\mathsf{item}(v));\ \mathsf{Continue}(1)\
  \mathbf{else}\ \mathbf{send}(o,\mathsf{end});\ \mathsf{Done}\\[4pt]
\mathsf{Value}(v)(t\to o):\\
\quad\mathbf{on}\ t\ \{\ \_\Rightarrow\mathbf{send}(o,\mathsf{item}(v));\mathsf{Continue}
  \mid\mathsf{end}\Rightarrow\mathbf{send}(o,\mathsf{end});\mathsf{Done}\\
\qquad\qquad\ \mid\mathsf{round}\Rightarrow\mathbf{send}(o,\mathsf{round});\mathsf{Continue}\ \}\\[4pt]
\mathsf{Source}(f)(t\to o):\quad\text{as }\mathsf{Value}\text{, sending }\mathsf{item}(f())\\[4pt]
\mathsf{Stateless}(f)(i\to o):\quad\text{as }\mathsf{Value}\text{ on }i\text{, sending }\mathsf{item}(f(x))
  \text{ for the item } x\\[4pt]
\mathsf{Fanout}(i\to o_1,\dots,o_n):\quad\text{as }\mathsf{Value}\text{ on }i\text{, sending what it
  receives on every } o_j\\[4pt]
\mathsf{For}(i\to e,k),\ s_0=[\,]:\\
\quad\mathbf{if}\ s\neq[\,]\ \mathbf{then}\ \mathbf{send}(e,\mathsf{item}(\mathbf{head}\ s));\
  \mathsf{Iterate}(\mathbf{tail}\ s)\\
\quad\mathbf{else}\ \mathbf{on}\ i\ \{\ \ell\Rightarrow\mathbf{send}(k,\mathsf{item}(\mathbf{length}\ \ell));\
  \mathbf{if}\ \ell=[\,]\ \mathbf{then}\ \mathsf{Continue}([\,])\\
\qquad\qquad\qquad\qquad\mathbf{else}\ \mathbf{send}(e,\mathsf{item}(\mathbf{head}\ \ell));\
  \mathsf{Iterate}(\mathbf{tail}\ \ell)\\
\qquad\qquad\ \ \mid\mathsf{end}\Rightarrow\mathbf{send}(e,\mathsf{end});\mathbf{send}(k,\mathsf{end});\mathsf{Done}\\
\qquad\qquad\ \ \mid\mathsf{round}\Rightarrow\mathbf{send}(e,\mathsf{round});\mathbf{send}(k,\mathsf{round});
  \mathsf{Continue}([\,])\ \}\\[4pt]
\mathsf{Reuse}(c,k\to o),\ s_0=((),0):\\
\quad\mathbf{if}\ s=(x,n)\ \text{with}\ n\neq 0\ \mathbf{then}\ \mathbf{send}(o,\mathsf{item}(x));\
  \mathsf{Continue}(x,n-1)\\
\quad\mathbf{else}\ \mathbf{on}\ c\ \{\ x\Rightarrow\mathbf{let}\ n=\mathbf{snd}\ \mathbf{recv}(k)\ \mathbf{in}\\
\qquad\qquad\qquad\qquad\mathbf{if}\ n\neq 0\ \mathbf{then}\ \mathbf{send}(o,\mathsf{item}(x));\
  \mathsf{Continue}(x,n-1)\ \mathbf{else}\ \mathsf{Continue}(x,0)\\
\qquad\qquad\ \ \mid\mathsf{end}\Rightarrow\mathbf{recv}(k);\mathbf{send}(o,\mathsf{end});\mathsf{Done}\\
\qquad\qquad\ \ \mid\mathsf{round}\Rightarrow\mathbf{recv}(k);\mathbf{send}(o,\mathsf{round});\mathsf{Continue}((),0)\ \}\\[4pt]
\mathsf{StateInit}(i\to q),\ s_0=\bot:\\
\quad\mathbf{if}\ s=\bot\ \mathbf{then}\ \mathsf{Continue}(\mathbf{snd}\ \mathbf{recv}(i))\
  \mathbf{else}\ \mathbf{recv}(i);\ \mathbf{send}(q,\mathbf{newstate}(s));\ \mathsf{Done}\\[4pt]
\mathsf{Stateful}(f)(q,i\to o,q'),\ s_0=\bot:\\
\quad\mathbf{let}\ \ell=(\mathbf{if}\ s=\bot\ \mathbf{then}\ \mathbf{recv}(q)\ \mathbf{else}\ s)\ \mathbf{in}\\
\quad\mathbf{on}\ i\ \{\ x\Rightarrow\mathbf{send}(o,\mathsf{item}(\ell.f(x)));\mathsf{Continue}(\ell)\\
\qquad\quad\mid\mathsf{end}\Rightarrow\mathbf{send}(o,\mathsf{end});\mathbf{send}(q',\ell);\mathsf{Done}\\
\qquad\quad\mid\mathsf{round}\Rightarrow\mathbf{send}(o,\mathsf{round});\mathbf{send}(q',\ell);\mathsf{Continue}(\bot)\ \}\\[4pt]
\mathsf{Stateful}_0(f)(q,t\to o,q'):\quad\text{as }\mathsf{Stateful}\text{ on the trigger }t\text{, sending }
  \mathsf{item}(\ell.f())\\[4pt]
\mathsf{Target}(i),\ s_0=\bot:\quad
  \mathbf{if}\ s=\bot\ \mathbf{then}\ \mathsf{Continue}(\mathbf{snd}\ \mathbf{recv}(i))\
  \mathbf{else}\ \mathbf{recv}(i);\ \mathsf{Done}(s)
\end{array}
\]
}
Unless given, the initial node state $s_0$ is $()$.
$\mathsf{StateInit}$ and $\mathsf{Target}$ read the single item of their input and then its end marker;
the item of $\mathsf{Target}$ is the result of the graph.
$\mathsf{Const}$, $\mathsf{StateInit}$ and $\mathsf{Target}$ never run inside the body of a $\mathsf{Fix}$
loop, so they have no round case.

$\mathsf{Fix}(n_p)(c_s,c_f,\bar t_{\mathit{in}},\bar t_{\mathit{back}}\to a,r,\bar t_{\mathit{body}},
\bar t_{\mathit{out}})$, with $n_p$ channels in each of the four lists of state channels, holds the
tokens $\bar\ell$ of its states and goes through five phases; it starts in $\mathsf{Init}$.
Below, $\mathit{start}(\bar\ell,x)$ sends the tokens $\bar\ell$ on $\bar t_{\mathit{body}}$, then
$\mathsf{item}(x)$ and $\mathsf{round}$ on $a$, and returns $\mathsf{Iterate}(\mathsf{Running})$.
{\small
\[
\begin{array}{@{}l@{}}
\mathsf{Init}:\quad\bar\ell\leftarrow\mathbf{recv}(\bar t_{\mathit{in}});\ \text{continue as }\mathsf{Idle}(\bar\ell)\\[4pt]
\mathsf{Idle}(\bar\ell):\quad\mathbf{on}\ c_s\ \{\ x\Rightarrow\mathit{start}(\bar\ell,x)\\
\qquad\mid\mathsf{end}\Rightarrow\mathbf{send}(a,\mathsf{end});\mathbf{send}(r,\mathsf{end});
  \mathbf{send}(\bar t_{\mathit{out}},\bar\ell);\\
\qquad\qquad\qquad\ \mathbf{send}(\bar t_{\mathit{body}},\mathsf{end});\mathsf{Continue}(\mathsf{Closing})\\
\qquad\mid\mathsf{round}\Rightarrow\mathbf{send}(r,\mathsf{round});\mathbf{send}(\bar t_{\mathit{out}},\bar\ell);
  \mathsf{Continue}(\mathsf{Init})\ \}\\[4pt]
\mathsf{Running}:\quad\mathbf{on}\ c_f\ \{\ y\Rightarrow\mathsf{Continue}(\mathsf{Wait}(y))\ \}\\[4pt]
\mathsf{Wait}(y):\quad\mathbf{on}\ c_f\ \{\ \mathsf{round}\Rightarrow\bar\ell\leftarrow\mathbf{recv}(\bar t_{\mathit{back}});\
  \mathbf{case}\ y\ \{\ \mathsf{Continue}(x')\Rightarrow\mathit{start}(\bar\ell,x')\\
\qquad\qquad\qquad\qquad\qquad\qquad\qquad\qquad\qquad\ \,
  \mid\mathsf{Done}(v)\Rightarrow\mathbf{send}(r,\mathsf{item}(v));\mathsf{Continue}(\mathsf{Idle}(\bar\ell))\ \}\ \}\\[4pt]
\mathsf{Closing}:\quad\mathbf{on}\ c_f\ \{\ \mathsf{end}\Rightarrow\text{receive }\mathsf{end}\text{ from every }
  t\in\bar t_{\mathit{back}};\ \mathsf{Done}\ \}
\end{array}
\]
}
A result $y$ that is neither $\mathsf{Continue}(x')$ nor $\mathsf{Done}(v)$ raises $\mathsf{ub}$, as in
the source.
At the end of its seeds, $\mathsf{Fix}$ sends end markers on $\bar t_{\mathit{body}}$: a stateful node of
the body, waiting for a token, takes the end marker as its token and passes it on, which closes the
chains of state channels of the body; $\mathsf{Closing}$ collects them.
A round marker on $c_s$ ends a round of an enclosing loop: $\mathsf{Fix}$ passes it on $r$, sends its
tokens on $\bar t_{\mathit{out}}$, and returns to $\mathsf{Init}$ to receive the tokens of the next outer
round.

\subsubsection{Denotation of graph programs}\label{app:tgt-sem}
The values of graph programs are the values of the source and the nodes.
The denotation $\llbracket g\rrbracket:\itreeER{E_{\mathit{dfg}}}{\mathsf{val}}$ is given below; it raises
$\mathsf{ub}$ on $\mathbf{fst}$ or $\mathbf{snd}$ of a value that is not a pair, on a node kind applied to
a node, and on a loop group or an execution of something that is not a node:
\[
\begin{array}{@{}r@{\ }c@{\ }l@{}}
\llbracket v\rrbracket &=& \mathsf{Ret}\ v\qquad \llbracket x\rrbracket\ =\ \mathsf{ub}\\
\llbracket\mathbf{chan}\rrbracket &=& (c_s,c_r)\leftarrow\mathsf{chan};\ \mathsf{Ret}\ (c_s,c_r)\\
\llbracket\mathbf{fst}\ g\rrbracket &=& (a,\_)\leftarrow\llbracket g\rrbracket;\ \mathsf{Ret}\ a
\\
\llbracket\mathbf{let}\ x=g_1\ \mathbf{in}\ g_2\rrbracket &=& v\leftarrow\llbracket g_1\rrbracket;\
  \llbracket g_2[v/x]\rrbracket\\
\llbracket k(g_1,\dots,g_m)\rrbracket &=& v_1\leftarrow\llbracket g_1\rrbracket;\dots;
  v_m\leftarrow\llbracket g_m\rrbracket;\ \mathsf{node}(k(v_1,\dots,v_m))
\\
\llbracket\mathbf{loop}(g_1,\dots,g_m)\rrbracket &=& n_1\leftarrow\llbracket g_1\rrbracket;\dots;
  n_m\leftarrow\llbracket g_m\rrbracket;\ \mathsf{node}(\mathbf{loop}(n_1,\dots,n_m))
\\
\llbracket\mathbf{execute}(g_1,\dots,g_m)\rrbracket &=& n_1\leftarrow\llbracket g_1\rrbracket;\dots;
  n_m\leftarrow\llbracket g_m\rrbracket;\ \mathsf{execute}(n_1,\dots,n_m)
.
\end{array}
\]
Every executor handles the construction events in the same way: $\mathsf{chan}$ allocates a fresh channel
with an empty queue on the communication heap, and answers with the channel for both of its ends;
$\mathsf{node}(n)$ answers with the node $n$ itself.

\subsubsection{Executors}\label{app:executors}
All executors share the following definitions.
A \emph{run} of $\mathsf{execute}(\bar n)$ keeps the communication heap and, for every node, either its
node state or, once it has returned $\mathsf{Done}(v)$, its value $v$; the nodes start in their initial
node states, and a loop group is only a group of nodes.
A node \emph{can fire} when it is not done, every input it reads in its node state holds data, and, for a
$\mathsf{For}$ that reads its input, the first value there is a list item or a marker.
\emph{Firing} a node runs one activation of its operator, with the communication events handled on the
communication heap; afterwards, the executor raises $\mathsf{iter}$ if the activation started an
iteration: a $\mathsf{For}$ that sent an element, or a $\mathsf{Fix}$ that took a seed or received
$\mathsf{Continue}$.
The run ends when every node is done, and its result is the value of the last node, which the compiler
makes the $\mathsf{Target}$ node; if no node can fire before that, the run raises $\mathsf{ub}$.
After interpreting the communication and construction events, an executor turns
$\llbracket g\rrbracket$ into an \itree over $\mathsf{funCallE}+\mathsf{progE}+\mathsf{ubE}+
\mathsf{stepE}+\mathsf{iterE}$, the events of the source (\Cref{sec:source-sem}) and $\mathsf{step}$.

\emph{Schedulers.}
A scheduler raises one $\mathsf{step}$ per round, and fires one node per round.
The \emph{source-order} policy keeps the position $i$ of the run in the trace of the graph
(\Cref{app:trace}), together with the lists that the $\mathsf{For}$ nodes have taken and the rounds that
the $\mathsf{Fix}$ nodes have revealed, which determine the trace.
If the node of the control firing at position $i$ can make it, the policy fires that node and moves to
$i+1$; otherwise it fires the first node, in the order of the graph, that can fire and makes no control
firing.
The \emph{demonic} scheduler raises a choice event among all nodes that can fire, answered by the
environment; its handler in the program logic quantifies universally over the answer.
For the proofs, it keeps the same bookkeeping as the source-order policy, and it also counts the control
firings that every node has made ahead of the position $i$; before every choice, the position catches up
with the firings already made. The bookkeeping does not influence the choice.

\emph{Executors without a scheduler.}
The \emph{local} executor is the demonic scheduler without its bookkeeping: in every round, it raises a
choice event among the nodes that can fire.
The \emph{parallel} executor takes, in every round, a snapshot of the communication heap and a
\emph{batch}: going through the nodes that can fire in the order of the graph, it adds every node whose
inputs are disjoint from those of the nodes already in the batch, leaving out nodes that read a channel
twice; if only such nodes can fire, the batch is one of them.
Every member of the batch fires on the snapshot, recording how many values it received from every channel
and what it sent; then, on every channel, the executor drops the received values from the head of the
queue and appends the sent values, in batch order.
It raises one $\mathsf{step}$ per activation, so a round with $k$ members has the events of $k$ rounds of
the local executor, and makes no choice.

\subsection{Compiler}\label{app:compiler}

\subsubsection{Compilation rules}\label{app:compile-rules}
The compiler is a function in a monad $M(A)\triangleq\mathbb{N}\to(A\times\mathbb{N})_\bot$: the number
is a counter of fresh channel names, and $\bot$ is failure.
It fails when a use finds no unused copy of its variable or of the trigger in the context, when a stateful
call names a state that has no state channel in the context, when a \sfor body breaks the loop-body rule,
and when a \strfix body allocates a state; by \Cref{lem:compile-linear}, none of these happens for a
well-formed program.
\Cref{fig:compile-rules} gives the rules for the constructs with an argument and for loops; the calls
without argument are compiled like their trigger:
\[
\inferrule*[right=Source Call]
  {T=t\cdot T'}
  {\langle V,\Sigma,T\rangle\vdash f()\rightsquigarrow(\mathsf{Source}(f)(t\to o),\ o,\
   \langle V,\Sigma,T'\rangle)}
\]
\[
\inferrule*[right=Stateful$_0$ Call]
  {T=t\cdot T' \\ \Sigma(s)=q}
  {\langle V,\Sigma,T\rangle\vdash s.f()\rightsquigarrow(\mathsf{Stateful}_0(f)(q,t\to o,q'),\ o,\
   \langle V,\Sigma[s\mapsto q'],T'\rangle)}
\]
with $o$ and $q'$ fresh.
In \textsc{For}, the loop's value takes its copy of the outer trigger after the body is compiled, and in
\textsc{Let} and \textsc{LetState}, the binding of $x$ or $s$ in the resulting context is restored to
what it was before.
A program $e$ is compiled as in \Cref{sec:compiler-impl}: a fresh channel $t_0$ for the program's
trigger, $\mathsf{copies}(t_0,\mathsf{triggers}(e))=(\bar t,G_t)$, the judgment
$\langle\emptyset,\emptyset,\bar t\rangle\vdash e\rightsquigarrow(G,c,\_)$, and the node list
$\mathsf{Const}(())(\to t_0)\,;G_t\,;G\,;\mathsf{Target}(c)$, executed after allocating every channel
that the counter has named.

\subsubsection{Copies and triggers}\label{app:copies}
$\mathsf{copies}(c,n)$ is $(\varepsilon,\varepsilon)$ for $n=0$, $(c,\varepsilon)$ for $n=1$, and
$(\bar c,\mathsf{Fanout}(c\to\bar c))$ with $n$ fresh channels $\bar c$ otherwise.
$\mathsf{occurs}(x,e)$ says whether $x$ occurs free in $e$, where the body of a \strfix counts as closed:
\[
\mathsf{occurs}(x,\mathbf{trfix}\ y=e_i\ \mathbf{in}\ e_b)\triangleq\mathsf{occurs}(x,e_i),
\]
and the other cases are the usual ones.
The number of copies of a variable that an expression takes, and of the trigger, are, writing $[P]$ for
$1$ if $P$ holds and $0$ otherwise,
\[
\begin{array}{@{}r@{\ }c@{\ }l@{}}
\mathsf{uses}(x,y) &=& [x=y]\qquad \mathsf{uses}(x,v)\ =\ 0\\
\mathsf{uses}(x,\mathbf{let}\ y=e_1\ \mathbf{in}\ e_2) &=& \mathsf{uses}(x,e_1)+
  [y\neq x]\cdot\mathsf{uses}(x,e_2)\\
\mathsf{uses}(x,f(e))\ =\ \mathsf{uses}(x,s.f(e)) &=& \mathsf{uses}(x,e)\\
\mathsf{uses}(x,f())\ =\ \mathsf{uses}(x,s.f()) &=& 0\\
\mathsf{uses}(x,\mathbf{for}\ y\ \mathbf{in}\ e_i\ \mathbf{do}\ e_b) &=& \mathsf{uses}(x,e_i)+
  [y\neq x\wedge\mathsf{occurs}(x,e_b)]\\
\mathsf{uses}(x,\mathbf{letstate}\ s=e_i\ \mathbf{in}\ e_b) &=& \mathsf{uses}(x,e_i)+\mathsf{uses}(x,e_b)\\
\mathsf{uses}(x,\mathbf{trfix}\ y=e_i\ \mathbf{in}\ e_b) &=& \mathsf{uses}(x,e_i)\\[6pt]
\mathsf{triggers}(v)\ =\ \mathsf{triggers}(f())\ =\ \mathsf{triggers}(s.f()) &=& 1\qquad
  \mathsf{triggers}(x)\ =\ 0\\
\mathsf{triggers}(f(e))\ =\ \mathsf{triggers}(s.f(e)) &=& \mathsf{triggers}(e)\\
\mathsf{triggers}(\mathbf{let}\ x=e_1\ \mathbf{in}\ e_2) &=& \mathsf{triggers}(e_1)+\mathsf{triggers}(e_2)\\
\mathsf{triggers}(\mathbf{letstate}\ s=e_1\ \mathbf{in}\ e_2) &=& \mathsf{triggers}(e_1)+\mathsf{triggers}(e_2)\\
\mathsf{triggers}(\mathbf{for}\ x\ \mathbf{in}\ e_i\ \mathbf{do}\ e_b) &=& \mathsf{triggers}(e_i)+1\\
\mathsf{triggers}(\mathbf{trfix}\ x=e_i\ \mathbf{in}\ e_b) &=& \mathsf{triggers}(e_i).
\end{array}
\]
In \textsc{For}, $\mathsf{reuse}_{e_b}(V,k)$ takes the variables $y_1,\dots,y_m$ of $V$, other than the
loop variable, that occur in $e_b$; it takes one copy $c_{y_j}$ of each from $V$, leaving $V'$, makes $m$
copies $k_{y_1},\dots,k_{y_m}$ of the extents on $k$, and emits, for every $j$,
$\mathsf{Reuse}(c_{y_j},k_{y_j}\to c'_{y_j})$ with a fresh $c'_{y_j}$ and
$\mathsf{copies}(c'_{y_j},\mathsf{uses}(y_j,e_b))$, whose channels form $V_b(y_j)$.

\subsubsection{Well-formed programs}\label{app:wf}
The states called in an expression, with repetitions, are
$\mathsf{sf}(s.f(e))=s\cdot\mathsf{sf}(e)$, $\mathsf{sf}(s.f())=s$, and
$\mathsf{sf}(\mathbf{trfix}\ x=e_i\ \mathbf{in}\ e_b)=\mathsf{sf}(e_i)\cdot\mathsf{dedup}(\mathsf{sf}(e_b))$,
concatenating over subexpressions otherwise: the $\mathsf{Fix}$ of a \strfix is the single site of every
state that its body calls.
$\mathsf{allocates}(e)$ holds when $e$ contains a \sletstate.
A loop body $e_b$ follows the loop-body rule when $\neg\mathsf{allocates}(e_b)$ and $\mathsf{sf}(e_b)$ has
no repetition.
The judgment $X;S\vdash e\ \mathsf{wf}$, for the variables $X$ and the states $S$ in scope, is
\[
\begin{array}{@{}c@{}}
\inferrule{ }{X;S\vdash v\ \mathsf{wf}}
\qquad
\inferrule{x\in X}{X;S\vdash x\ \mathsf{wf}}
\qquad
\inferrule{X;S\vdash e_1\ \mathsf{wf} \\ X,x;S\vdash e_2\ \mathsf{wf}}
  {X;S\vdash\mathbf{let}\ x=e_1\ \mathbf{in}\ e_2\ \mathsf{wf}}
\\[12pt]
\inferrule{X;S\vdash e\ \mathsf{wf}}{X;S\vdash f(e)\ \mathsf{wf}}
\qquad
\inferrule{s\in S \\ X;S\vdash e\ \mathsf{wf}}{X;S\vdash s.f(e)\ \mathsf{wf}}
\\[12pt]
\inferrule{X;S\vdash e_i\ \mathsf{wf} \\ X,x;S\vdash e_b\ \mathsf{wf} \\\\
  e_b\text{ follows the loop-body rule}}
  {X;S\vdash\mathbf{for}\ x\ \mathbf{in}\ e_i\ \mathbf{do}\ e_b\ \mathsf{wf}}
\qquad
\inferrule{X;S\vdash e_i\ \mathsf{wf} \\ X;S,s\vdash e_b\ \mathsf{wf}}
  {X;S\vdash\mathbf{letstate}\ s=e_i\ \mathbf{in}\ e_b\ \mathsf{wf}}
\\[12pt]
\inferrule{X;S\vdash e_i\ \mathsf{wf} \\ x;S\vdash e_b\ \mathsf{wf} \\ \neg\mathsf{allocates}(e_b)}
  {X;S\vdash\mathbf{trfix}\ x=e_i\ \mathbf{in}\ e_b\ \mathsf{wf}}
\end{array}
\]
with $f()$ and $s.f()$ as $f(e)$ and $s.f(e)$ without the premise on $e$.
A program $e$ is well formed when $\emptyset;\emptyset\vdash e\ \mathsf{wf}$.

\subsubsection{Linearity}\label{app:linear}
The \emph{emission order} of a compiled graph is the order of its node list after replacing every loop
group by its nodes.
The \emph{back inputs} of a $\mathsf{Fix}$ are its inputs $c_f$ and $\Sigma_b(\bar s)$ of rule
\textsc{TrFix}; all other inputs of every node are \emph{forward inputs}.
\Cref{lem:compile-linear} is the conjunction of three results, for a compiled program $g$ with node list
$\bar n$ in emission order:
\begin{enumerate}
\item the compiler succeeds on every well-formed program;
\item no channel is an output of two nodes of $\bar n$, nor an input of two nodes, and every forward input
of a node is an output of a node that comes before it in $\bar n$;
\item every back input of a node is an output of some node of $\bar n$.
\end{enumerate}

\subsection{Graph Logic and \texorpdfstring{$\mathrm{WP}_{\mathsf{t}}$}{WPt}}\label{app:graph-logic}

\subsubsection{Handlers of graph programs}\label{app:graph-handlers}
Channels are described by four kinds of resources: the persistent fact $c[n]=v$, that the
$n$-th value sent on $c$ is $v$; the exclusive send end $c\rightarrowtail_{\mathsf{s}}n$ and receive end
$c\rightarrowtail_{\mathsf{r}}n$, which have sent and received $n$ values; and a persistent tag
$c\rightsquigarrow\tau$, which names the static edge that $c$ belongs to and the instance of its scope.
For every tag $\tau$, a protocol $\Psi_\tau(n,v)$ says what the $n$-th value of a channel with tag $\tau$
carries: sending gives it away, receiving gets it.
The handlers of construction and communication are
{\small
\[
\begin{array}{@{}r@{\ }c@{\ }l@{}}
H_{\mathit{constr}}(\mathsf{chan},\Phi) &\triangleq& \forall c.\ c\rightarrowtail_{\mathsf{s}}0\ast
  c\rightarrowtail_{\mathsf{r}}0\ast c\rightsquigarrow{?}\mathrel{-\!\!*}\Phi(c,c)\qquad
H_{\mathit{constr}}(\mathsf{node}(n),\Phi)\ \triangleq\ \Phi(n)\\[2pt]
H_{\mathit{comm}}(\mathsf{send}(c,v),\Phi) &\triangleq& \exists n,\tau.\ c\rightarrowtail_{\mathsf{s}}n\ast
  c\rightsquigarrow\tau\ast\Psi_\tau(n,v)\ast(c\rightarrowtail_{\mathsf{s}}n{+}1\mathrel{-\!\!*}
  c[n]=v\mathrel{-\!\!*}\Phi())\\[2pt]
H_{\mathit{comm}}(\mathsf{recv}(c),\Phi) &\triangleq& \exists n,\tau,w.\ c\rightarrowtail_{\mathsf{r}}n\ast
  c[n]=w\ast c\rightsquigarrow\tau\ast(c\rightarrowtail_{\mathsf{r}}n{+}1\mathrel{-\!\!*}
  \Psi_\tau(n,w)\mathrel{-\!\!*}\Phi(w)),
\end{array}
\]
}
where $c\rightsquigarrow{?}$ is a one-shot token that the builder of the channel's writer turns into its
tag, since a graph allocates its channels before its nodes.
A receive needs the fact of the value it receives, which the executor provides from the queue when it
fires the node.
$H_{\mathit{step}}(\mathsf{step},\Phi)$ is $\triangleright\Phi()$ in the step mode $\mathsf{Later}$, which
the theorems of \Cref{sec:verification} assume, and $\Phi()$ otherwise; $H_{\mathit{ub}}$ is as in the
source.
The handler of iterations $H_{\mathit{iter}}$ is a parameter, shared with the source.

\subsubsection{The trace}\label{app:trace}
A graph is executed as a \emph{process tree} $T$, whose leaves are its nodes and whose inner nodes are
its loop groups; a node is named by its path $p$ in $T$.
A \emph{position} $\pi=(p,j)$ is a node with the number $j$ of its control firings before: its calls for
a call node, its activations for a loop head.
The control firings are
\[
\mathsf{ECall}(\pi)\quad
\mathsf{ETake}(\pi,k)\quad
\mathsf{EElem}(\pi)\quad
\mathsf{ESeed}(\pi,k)\quad
\mathsf{EDec}(\pi,k,j):
\]
a call; the $k$-th take of a list by a $\mathsf{For}$; the sending of a further element; the take of the
$k$-th seed by a $\mathsf{Fix}$; and its decision after the $j$-th round of that seed.
Given a function $L$ that maps every $\mathsf{For}$ and index $k$ to the $k$-th list it takes, and every
$\mathsf{Fix}$ and seed index $k$ to the arguments of the rounds of that seed, the \emph{trace}
$\mathsf{trace}(L,T)$ is the list of the control firings of $T$ in the order of the source program.
It is defined by recursion on $T$ and follows the structure that the compiler gives it: statements in
order, one segment of a loop body per element of the list taken, and one segment of a tail-recursive body
per round.
A control firing $e$ is an \emph{iteration} when it is the take of a non-empty list or of a seed, a further
element, or a decision that starts another round; it is a \emph{skip} when it is the take of an empty list
or a decision that ends the rounds of its seed.

\subsubsection{The graph specification}\label{app:wpt}
The ghost state of $\mathrm{WP}_{\mathsf{t}}$ has, for every node $p$, an exclusive counter
$\mathsf{cnt}(p,j)$ of its control firings and an exclusive counter $\mathsf{itc}(p,j)$ of its iterations,
with authorities $\mathsf{cnt}_\bullet(M)$ and $\mathsf{itc}_\bullet(M')$; a persistent record
$\mathsf{ans}(\pi,r)$ of the answer $r$ of the call at $\pi$; and a persistent permission $\mathsf{go}(\pi)$
for the call at $\pi$.
The call handler of the node $p$, given persistent certificates $\mathit{Arg}(\pi,\iota)$ of what the call
at $\pi$ is about ($\iota$ names the function, the arguments and, for a stateful call, the state's value
before), is
\[
\begin{array}{@{}r@{\ }c@{\ }l@{}}
H^{p}_{\mathit{call}}(\mathsf{slcall}(f,\bar v),\Phi) &\triangleq& \exists j.\ \mathsf{cnt}(p,j)\ast
  \mathsf{go}(p,j)\ast\mathit{Arg}((p,j),f\,\bar v)\ast{}\\
  && \quad\forall r.\ \mathsf{spec}_f(\bar v,r)\mathrel{-\!\!*}\mathsf{ans}((p,j),r)\mathrel{-\!\!*}
  \mathsf{cnt}(p,j{+}1)\mathrel{-\!\!*}\Phi(r),
\end{array}
\]
and likewise for $\mathsf{sfcall}$, with the points-to of the state as in $H_{\mathit{fun}}$.
The specification of the node $n$ at $p$, with an invariant $\mathit{Inv}$ on its node state and a
postcondition $\Psi$ on its value, is
\[
\mathit{Inv}(s_0)\ast\Box\,\forall s.\ \mathit{Inv}(s)\mathrel{-\!\!*}\mathsf{ready}(p,n,s,
  \mathrm{wp}_{\mathit{act}}(n,s,\lambda a.\ \mathit{post}(a)\ast\mathit{oblig}(a))),
\]
where $\mathsf{ready}$ gives the activation the facts of the values it receives, its counters, and, for a
$\mathsf{For}$, that it reads a list item or a marker; $\mathrm{wp}_{\mathit{act}}$ is the weakest
precondition of one activation under $H_{\mathit{comm}}$, $H^{p}_{\mathit{call}}$ and the handlers of the
program heap, undefined behaviour and steps; $\mathit{post}(a)$ is $\mathit{Inv}(s')$ if
$a=\mathsf{Continue}(s')$ or $\mathsf{Iterate}(s')$, and $\Psi(v)$ if $a=\mathsf{Done}(v)$; and
$\mathit{oblig}(a)$ says that an activation that reaches a call makes it, and that the take of a list or a
seed and the decision of a $\mathsf{Fix}$ come with persistent certificates $\mathit{Tk}$ and
$\mathit{Dk}$ of the value they read.

$\WPt{T}{\Phi}$, for the process tree $T$ of the nodes $\bar n$ of \Cref{sec:verification}, holds when there exist predicates $\mathsf{Ctl}$ on positions, $\mathit{Arg}$,
$\mathit{Tk}$ and $\mathit{Dk}$, all three persistent, and $\mathit{Inv}_p$ and $\Psi_p$ for every node,
such that:
\begin{enumerate}[label=(\roman*)]
\item $\mathsf{Ctl}$ holds at the first position of the trace;
\item for every $L$ and every position $i$ of $\mathsf{trace}(L,T)$ whose firing is a call at $\pi$
  followed by the position $\pi'$: given the facts of the takes up to $i$, $\mathit{Arg}$ of the call,
  $\mathsf{Ctl}(\pi)$, an answer $r$ allowed by $\mathsf{spec}_f$ and $\mathsf{ans}(\pi,r)$, a ghost update
  gives $\mathsf{Ctl}(\pi')$;
\item for every iteration at $\pi$ followed by $\pi'$, with the same facts, $\mathsf{Ctl}(\pi)$ gives,
  after a ghost update, $H_{\mathit{iter}}(\mathsf{iter},\lambda\_.\,\mathsf{Ctl}(\pi'))$; for every skip,
  it gives $\mathsf{Ctl}(\pi')$;
\item every node meets its specification, from its counters at $0$;
\item for every reachable state of the run, given the invariant of the channels, $\mathsf{Ctl}$ at the
  position $i$, the invariants and postconditions of the nodes, and the authorities of the counters, at
  the counts of the first $i$ firings except that a call node or a loop head other than the one of the
  firing at $i$ may be ahead, a ghost update gives all of them back and the fact that either every node
  is done and the trace has no firing left, or the node of the firing at $i$ can make it, or some node can
  fire without a control firing;
\item $\mathsf{Ctl}$ past the last position and the postconditions $\Psi_p$ of the nodes, applied to their
  values, give $\Phi$ of the value of the last node.
\end{enumerate}
The proof of \Cref{lem:wp-pres} takes for $\mathsf{Ctl}$ the greatest fixpoint of the conjunction of the
progress obligation (v) at a position and the transitions (ii) and (iii) out of it, with (vi) at the end.
It then establishes $\mathsf{Ctl}$ at the start of the trace by the following lemma.

\begin{lemma}[Control lemma]\label{lem:control}
Let $e$ compile, in a context $C$, to the nodes $\bar n$, whose segment of the trace starts at the position
$\pi$.
Assume that the variables of $C$ are known (their channels carry the knowledge of their source values),
that the states of $C$ are related to their state channels on the ghost instance of the heap, and that
$\wpi{H_{\mathit{src}}}{x\leftarrow\llbracket e\rrbracket;\,k\,x}{\Phi}$ holds.
Assume moreover that for every value $v$ of $e$, the knowledge that the result channel of $e$ carries
$v$, the states related again, and $\wpi{H_{\mathit{src}}}{k\,v}{\Phi}$ give $\mathsf{Ctl}$ at the
position after the segment.
Then $\mathsf{Ctl}(\pi)$ holds.
\end{lemma}
The lemma is proved by induction on $e$, one case per compilation rule; inside a loop body and a round of
a tail recursion it is stated per row, and the rounds are proved by induction on the source's weakest
precondition of the loop, a least fixpoint.
At each position, the proof gives progress and, after every transition, $\mathsf{Ctl}$ at the next
position.
At a call, the node's certificate rests on the persistent knowledge that channels carry about their items:
it makes the graph's arguments, and for a stateful call the value of the state's token, the source's, so
the source's next event is the same call.
The source's heap operations run on a separate ghost instance of the heap, which its weakest precondition
allows, since it holds for every instance; an allowed answer, once recorded, continues both sides.
At an iteration, the source also makes an iteration, under the same $H_{\mathit{iter}}$, and the
certificates of $\mathsf{For}$ and $\mathsf{Fix}$ show that they take the source's lists and decide as its
rounds do; only these decisions reveal how many rounds a tail recursion makes.
Progress holds because an item or a marker is available without a control firing whenever the inputs of
its writer are, down to control firings already made and the program's trigger.
In particular, a $\mathsf{Fix}$ is at its control firing or can read its back edges, and after the last
position the first node in emission order that is not done can fire.
At the end, the knowledge of the program's result channel makes the graph's result the source's value, for
which the source's weakest precondition has become $\Phi$.

\subsubsection{Weakest preconditions of executors}\label{app:wpx}
Every executor $x$ turns a graph program $g$ into an \itree $\mathsf{run}_x(g)$ over the events of the
source, $\mathsf{funCallE}+\mathsf{progE}+\mathsf{ubE}+\mathsf{stepE}+\mathsf{iterE}$
(\Cref{app:executors}), with a choice event for the demonic scheduler and the local executor.
With $H_{\mathit{node}}=H_{\mathit{fun}}\oplus H_{\mathit{heap}}\oplus H_{\mathit{ub}}\oplus
H_{\mathit{step}}$ and $H_{\forall}$ the handler of choice that quantifies universally over the answer,
\[
\begin{array}{@{}r@{\ }c@{\ }l@{}}
\WPx{order}{g}{\Phi} &\triangleq& \wpi{H_{\mathit{node}}\oplus H_{\mathit{iter}}}{\mathsf{run}_{\mathit{order}}(g)}{\Phi}\\
\WPx{x}{g}{\Phi} &\triangleq& \wpi{H_{\forall}\oplus H_{\mathit{node}}\oplus H^{\mathit{free}}_{\mathit{iter}}}{\mathsf{run}_x(g)}{\Phi}
  \qquad x\in\{\mathit{demonic},\mathit{local}\}\\
\WPx{par}{g}{\Phi} &\triangleq& \wpi{H_{\mathit{node}}\oplus H^{\mathit{free}}_{\mathit{iter}}}{\mathsf{run}_{\mathit{par}}(g)}{\Phi},
\end{array}
\]
where $H^{\mathit{free}}_{\mathit{iter}}(\mathsf{iter},\Psi)\triangleq\Psi(())$ is the handler of
unguarded iterations.

\subsubsection{Theorems}\label{app:theorems}
For a well-formed program $e$ that compiles to $g$, every postcondition $\Phi$, and the step mode
$\mathsf{Later}$, the theorems of \Cref{sec:verification} state
\[
\begin{array}{@{}r@{\ }c@{\ }l@{\qquad}l@{}}
\WPsrc{e}{\Phi} &\vdash& \WPg{g}{\Phi^{\uparrow}} & \text{for every } H_{\mathit{iter}}
  \text{ (\Cref{lem:wp-pres}, every step mode)}\\
\WPsrc{e}{\Phi} &\vdash& \WPx{order}{g}{\Phi^{\uparrow}} & \text{for every } H_{\mathit{iter}}\
  \text{(\Cref{thm:wp-pres-order})}\\
\WPsrctot{e}{\Phi} &\vdash& \WPx{x}{g}{\Phi^{\uparrow}} & x\in\{\mathit{demonic},\mathit{local},\mathit{par}\}\
  \text{(\Cref{thm:wp-pres-sched,thm:wp-pres-distr})},
\end{array}
\]
where $\WPsrc{e}{\Phi}$ quantifies over the instances of the ghost state of the program heap and the
executor's weakest preconditions are at one instance.
The proofs of \Cref{thm:wp-pres-distr} use two relations between executors: the runs of the local
executor and of the demonic scheduler are equal up to $\mathsf{Tau}$ steps, and the run of the parallel
executor on a compiled graph is a \emph{resolution} of the local executor's choices: it is obtained from it
by answering every choice event, up to $\mathsf{Tau}$ steps.

\emph{Adequacy.}
An implementation $\mathit{ctx}$ of the functions meets their specifications when every call that it
answers satisfies $\mathsf{spec}_f$, proved in the program logic.
Let the source be verified: for such a $\mathit{ctx}$, $\WPsrctot{e}{\lambda v.\,\varphi(v)}$ for a pure
postcondition $\varphi$.
Run from the empty program heap, with the calls answered by $\mathit{ctx}$ and a budget of $n$ steps,
every run of the compiled graph, by the source-order scheduler, by every resolution of the choices of
the demonic scheduler or the local executor, and by the parallel executor, is equivalent up to
$\mathsf{Tau}$ steps to a run that returns, and its result is either a graph value $v'$ with
$\varphi(v')$ or the exhaustion of the budget; it is never undefined behaviour.

\end{document}